\documentclass{aa}  

\usepackage{graphicx}
\usepackage[colorlinks,linkcolor=blue,anchorcolor=blue,citecolor=blue]{hyperref}
\usepackage{natbib}
\usepackage{txfonts}
\usepackage[fleqn]{amsmath}
\newcommand{\beq}{\begin{equation}}
\newcommand{\eeq}{\end{equation}}
\newcommand{\bea}{\begin{eqnarray}}
\newcommand{\eea}{\end{eqnarray}}
\newcommand{\rem}[1]{ }

\def\cm{cm$^{-2}$}
\def\ergs{erg~s$^{-1}$}

\def\nh{{$N_{\rm H}$}}
\def\nhism{{$N_{\rm H,ISM}$}}

\def\chandra{{\it Chandra}}

\def\msun{M_{\odot}}
\def\msunpc{$M_{\odot}$~pc$^{-2}$}
\def\lsun{L_{\odot}}
\def\cii{[C\,{\textsc{ii}}]}
\def\ciii{C\,{\textsc{iii}}]}
\def\heii{He\,{\textsc{ii}}}
\def\civ{C\,{\sc iv}}
\def\nev{[Ne\,{\sc v}]}
\def\oiii{[O\,{\sc iii}]}
\def\nii{[N\,{\sc ii}]}

\def\lesssim{\mathrel{\hbox{\rlap{\hbox{\lower3pt\hbox{$\sim$}}}\hbox{\raise2pt\hbox{$<$}}}}}
\def\gtrsim{\mathrel{\hbox{\rlap{\hbox{\lower3pt\hbox{$\sim$}}}\hbox{\raise2pt\hbox{$>$}}}}}

\begin{document} 

\title{Supermassive Black Holes at High Redshift are Expected to be Obscured by their Massive Host Galaxies' Inter Stellar Medium}

\titlerunning{High-z SMBHs are expected to be obscured by their hosts' ISM}
\authorrunning{R. Gilli et al.}

   \author{
   R.~Gilli\inst{1},
   C.~Norman\inst{2,3},
   F.~Calura\inst{1},
   F.~Vito\inst{1},
   R.~Decarli\inst{1},
   S.~Marchesi\inst{1,4},
   K.~Iwasawa\inst{5,6},
   A.~Comastri\inst{1},
   G.~Lanzuisi\inst{1},
   F.~Pozzi\inst{7,1},
   Q.~D'Amato\inst{8,9},
   C.~Vignali\inst{7,1},
   M.~Brusa\inst{7,1},
   M.~Mignoli\inst{1},
   and P. Cox\inst{10}
          }

   \institute{INAF -- Osservatorio di Astrofisica e Scienza dello Spazio di Bologna, Via P. Gobetti 93/3, 40129 Bologna, Italy\\
              \email{roberto.gilli@inaf.it}
         \and
         Department of Physics and Astronomy, Johns Hopkins University, Baltimore, MD 21218, USA
         \and
         Space Telescope Science Institute, 3700 San Martin Drive, Baltimore, MD 21218, USA
         \and
         Department of Physics and Astronomy, Clemson University,  Kinard Lab of Physics, Clemson, SC 29634, USA
         \and
         Institut de Ci\`encies del Cosmos (ICCUB), Universitat de Barcelona (IEEC-UB), Mart\'i i Franqu\`es, 1, 08028 Barcelona, Spain
         \and
         ICREA, Pg. Llu\'is Companys 23, 08010 Barcelona, Spain
         \and
         Dipartimento di Fisica e Astronomia, Universit\`a degli Studi di Bologna, Via P. Gobetti 93/2, 40129 Bologna, Italy
	\and
	SISSA, Via Bonomea 265, 34136 Trieste, Italy
	\and
	INAF -- Istituto di Radioastronomia, Via P. Gobetti 101, 40129 Bologna, Italy
         \and
 	 Institut d’Astrophysique de Paris, 98bis Boulevard Arago, F-75014 Paris, France
         }

  \date{Received; accepted}
 
  \abstract
{We combine results from deep ALMA observations of massive ($M_*>10^{10}\;M_{\odot}$) galaxies at different redshifts to show that the column density of their inter stellar medium (ISM) rapidly increases towards early cosmic epochs. Our analysis includes objects from the ASPECS and ALPINE large programs, as well as individual observations of $z\sim 6$ QSO hosts. When accounting for non-detections and correcting for selection effects, we find that the median surface density of the ISM of the massive galaxy population evolves as  $\sim(1+z)^{3.3}$. This means that the ISM column density towards the nucleus of a $z>3$ galaxy is typically $>100$ times larger than locally, and it may reach values as high as Compton-thick at $z\gtrsim6$. Remarkably, the median ISM column density is of the same order of what is measured from X-ray observations of large AGN samples already at $z\gtrsim2$.

We develop a simple analytic model for the spatial distribution of ISM clouds within galaxies, and estimate the total covering factor towards active nuclei when obscuration by ISM clouds on the host scale is added to that of pc-scale circumnuclear material (the so-called 'torus'). The model includes clouds with a distribution of sizes, masses, and surface densities, and also allows for an evolution of the characteristic cloud surface density with redshift, $\Sigma_{c,*}\propto(1+z)^\gamma$. We show that, for $\gamma=2$, such a model successfully reproduces the increase of the obscured AGN fraction with redshift that is commonly observed in deep X-ray surveys, both when different absorption thresholds and AGN luminosities are considered. 

Our results suggest that 80-90\% of supermassive black holes in the early Universe ($z>6-8$) are hidden to our view, primarily by the ISM in their hosts. We finally discuss the implications of our results and how they can be tested observationally with current and forthcoming facilities (e.g. VLT, E-ELT, ALMA, JWST) and with next-generation X-ray imaging satellites. By extrapolating the observed X-ray nebulae around local AGN to the environments of SMBHs at high redshifts, we find $\lesssim 1$'' nebulae impose stringent design constraints on the spatial resolution of any future X-ray imaging great observatory in the coming decades.
}


   \maketitle

\section{Introduction}

Deep X-ray observations of the Universe have shown that most supermassive black holes (SMBHs) at galaxy centers grow hidden by dust and gas (see e.g. \citealt{hickox18} for a review) and that the incidence of obscuration in Active Galactic Nuclei (AGN) strongly increases with redshift \citep{Lafranca05, tu06, liu17, lanzuisi18, vito18, iwasawa20}.
This is especially true for intrinsically luminous systems, as the fraction of QSOs with $L_{bol}\gtrsim10^{45}$ \ergs\ ($L_X>10^{44}$ \ergs) that are obscured by hydrogen column densities of $N_H>10^{23}$ \cm\  rises from 10-20\% in the local Universe to $\sim$80\% at $z\sim4$, when the Universe was only 1.5 Gyr old \citep{vito18}. 
The origin of this trend is not yet understood. 

Indirect arguments suggest that the observed increase may not be linked to the incidence of small-scale obscuration at different cosmic epochs, i.e. in the physical and geometric properties of the circumnuclear medium surrounding the nucleus that is often associated with the torus of dusty gas postulated by the Unified Models (see \citealt{netzer15} and \citealt{ramos17} for recent reviews). Remarkably, the broad-band Spectral Energy Distributions (SEDs) of UV-bright QSOs look the same across cosmic time \citep{richards06,leipski14,bianchini19}. These SED's include a prominent rest-frame near-IR bump produced by dust heated to $T\sim 1000-1500$~K by the QSO UV radiation field and distributed on a few pc scales around the nucleus \citep{netzer15}. The QSO near-IR to UV luminosity ratio is a measure of the dust covering factor, which is of the order of 30-50\% \citep{lusso13}. The self-similarity of QSO SEDs across cosmic time suggests that this ratio is constant with redshift, and so consequently is the covering factor of the circumnuclear dusty gas.

Other arguments suggest instead that the increased obscuration with redshift might occur on the host galaxy scale ($\lesssim$~kpc). The physical properties of the interstellar medium (ISM) in galaxies change rapidly with cosmic time. Empirical trends are being measured in several samples of star forming galaxies spanning a broad redshift range, where the amount of cold gas, which contains most of the ISM mass, is seen to rapidly grow towards early cosmic time \citep{scoville17, tacconi18, aravena20}, possibly as fast as $M_{gas}\sim(1+z)^2$\citep{scoville17}. 
At the same time, galaxies have been observed to become progressively smaller at high redshifts, with typical size decreasing as $r\sim(1+z)^{-1}$ \citep{allen17}. It then seems unavoidable that the ISM density rapidly grows towards early cosmic epochs, possibly as fast as $\rho_{gas}\sim M_{gas}/r^3\sim (1+z)^5$. Similarly, the column density of the ISM, $N_{H,gas}\sim M_{gas}/r^2$, may grow as fast as $\sim (1+z)^4$, reaching values of $10^{23}$\cm\ or even Compton-thick (\nh$>1/\sigma_T\sim 1.5\times 10^{24}$\cm, where $\sigma_T$ is the Thomson cross-section) in the distant Universe, where large, $M_{gas}\sim10^{10}\msun$ ISM reservoirs have been observed in compact galaxies of characteristic size 1-2 kpc (see e.g. \citealt{hd20} for a recent review on distant, star-forming galaxies). In massive galaxies, the ISM is metal-rich \citep{lequeux79,tremonti04,maiolino08} and such column densities would absorb the X-ray nuclear radiation very effectively, adding to the obscuration produced by the small-scale circumnuclear medium. By using ALMA mm observations of a $z=4.75$ Compton-thick QSO in the CDFS, \citet{gilli14} found that the ISM column density measured in this source was comparable with the absorption to the nucleus measured from the X-ray spectrum. The same analysis was then expanded to a small sample of distant, obscured AGN in the redshift range $z\sim2.5-4.7$ by \citet{circosta19} and \citet{damato20cdfs}, who also found a good correspondence between the level of nuclear absorption measured in the X-rays ($\sim10^{23-24}$\cm) and the ISM column density measured through ALMA data. 

Numerical simulations also suggest that the ISM column density in distant galaxies is substantial \citep{trebitsch19,ni20,lupi22}. For example, \citet{ni20} found that, in the early Universe, $z>7$,  the ISM column density and covering factor of galaxy nuclei are so large that 99\% of early SMBHs in their simulated galaxies grow hidden by gas column densities of \nh$>10^{23}$\cm.

This work focuses on galaxies with stellar masses $M_*>10^{10}\;M_{\odot}$, as: i) they are the hosts of active SMBHs detectable in deep X-ray surveys up to the highest redshifts \citep{lanzuisi17,yang17}; ii) despite the overall decrease of galaxy metallicity with redshift, in such massive systems the ISM metallicity remains close to solar even at $z\sim2$ \citep{maiolino08,mama19} and drops to only $\lesssim0.5\times$ solar at $z>4$ according to simulations \citep{torrey19}. We recall that, in the literature, solar metallicities and abundances are usually assumed to convert the column density of metals estimated through X-ray spectroscopy of large AGN samples into a total gas (hydrogen) column density \citep{tozzi06,lanzuisi17,liu17,vito18}. In massive galaxies, the comparison between the ISM-based and X-ray-based column densities is then fair.
We use sensitive ALMA observations to determine the column density towards the nuclei of galaxies in samples at different redshifts. We then derive the cosmological evolution of the average ISM column density towards galaxy nuclei and of the fraction of ISM-obscured AGN, and compare with observations from deep X-ray surveys.

The paper is organized as follows. In Section \ref{samples} we present the ALMA galaxy samples used in this work. In Section~\ref{methods} we describe the methodology adopted to derive gas masses and sizes for the galaxies in our samples. In Section~\ref{results} we present our main results on the cosmic evolution of the ISM column density and work out a simple analytic model to quantify the cosmic evolution of the ISM-obscured AGN fractions. In Section~\ref{discuss} we discuss our findings, add the contribution of a small-scale torus to our ISM obscuration model, and compare our results with the fractions of obscured AGN observed in different X-ray samples. We also compare our findings with expectations from numerical simulations, and finally present some possible observational diagnostics to test the proposed ISM-obscuration scenario. We draw
our conclusions in Section~\ref{conclusions} . We also note in the Discussion and Conclusions significant design constraints on future X-ray imaging Great Observatories that are consequences of this work.

A concordance cosmology with $H_0=70$ km~s$^{-1}$~Mpc$^{-1}$, $\Omega_m=0.3$ , and $\Omega_\Lambda=0.7$, in agreement within the errors with the Planck 2015 results \citep{planck16} is used throughout this paper. All stellar masses in the considered galaxy samples were originally computed using a Chabrier IMF \citep{chabrier03}.

\begin{table*}
\centering
\caption{\label{tab:sample}Summary of the galaxy samples with $M_*>10^{10}\msun$ and reliable size estimate used in this work (see text for details).}
\begin{tabular}{rccrrrrl}
\hline \hline
Parent sample& Redshift& Band& Wav.& $rms$& $N^{det}_{cont.}$& $N^{det}_{\cii}$& Ref.\\
& & & (mm)&($\mu$Jy)& & & \\
(1)& (2)& (3)& (4)& (5)& (6)& (7)& (8)\\  
\hline
ASPECS&       0.4--2.8&  6&             1.2&        10&  32&   --&  \citet{aravena20}\\
COSMOS&      3.2--3.6&  6&             1.3&   10-50&    6&   --&  \citet{suzuki21}\\
ALPINE&         4.1--5.9&  7&   0.85,1.05&   28-50&    7&   9&  \citet{fujimoto20}\\
High-Lum QSOs& 5.7--7.6&  6&             1.2& 50-200&  27&  27& \citet{venemans20}\\
SHELLQs&        5.9--6.4&  6&       1.1,1.2&   10-20&    6&   7&  \citet{izumi18,izumi19}\cr
\hline                               
\end{tabular}
\tablefoot{Columns: (1) Parent survey or sample; (2) redshift range; (3) ALMA detection band; (4) observed wavelength; (5) continuum $rms$; (6) number of continuum detections; (7) number of \cii\ detections; (8) references.} 
\end{table*}

\section{Galaxy samples}\label{samples}

We searched for samples of galaxies with $M_*>10^{10}\msun$ featuring deep ALMA FIR/mm observations. We require a continuum $rms$ smaller than a few tens of $\mu$Jy
in band 6  or 7 ($\sim 1.2$~mm and $\sim 850 \mu$m, respectively). These bands are commonly used for dust observations as they maximize the ALMA observing efficiency by targeting wavelengths close to the redshifted FIR SED peak of distant galaxies. The required $rms$ allows us to probe dust masses down to $\approx 10^7\msun$ at all redshifts (see Fig.~\ref{mdz}), corresponding to ISM masses down to a few $\times10^9\msun$ (see next Sections), and to detect $>50\%$ of massive galaxies. A large FIR/mm detection rate is indeed key to perform any reliable correction for those systems missed by ALMA, likely poorer in gas, and hence to derive a sensible value for the average ISM density of the entire massive galaxy population. For objects at $z>4$, the [C \textsc{ii}]157.74~$\mu$m fine structure emission line is redshifted at $\gtrsim 800\, \mu$m, where the atmosphere is relatively transparent and allows effective observations with ALMA. As the \cii\ line is a major coolant for cold ($T\lesssim300$~K) gas \citep{wolfire03} and one of the brightest emission lines in star-forming galaxies \citep{stacey91,stacey10}, we will use it to probe the ISM content of distant galaxies in addition to dust emission. The main samples considered in this work are described below and summarized in Table~\ref{tab:sample} sorted by increasing redshift. The source selection criteria at the basis of these samples are admittedly not homogeneous, nonetheless we compiled a large, ``best-effort'' collection of galaxies with sufficiently good data to determine their ISM column density. In Section \ref{biases}, we try to correct {\it a posteriori} for the main selection biases and, in turn, draw general conclusions for the whole population of massive galaxies.

\subsection{ASPECS}\label{aspecs}
 
The ALMA Spectroscopic Survey (ASPECS) Large Program \citep{decarli19,aravena20,walter20} is a 4.6 arcmin$^2$ blank-field survey at the center of the 7Ms CDFS \citep{luo17}. The ASPECS area falls within the 11 arcmin$^2$ covered by the Hubble Ultra Deep Field (HUDF, \citealt{beckwith06}), and roughly overlaps with the Hubble eXtremely Deep Field (XDF), i.e. the region of the HUDF with the deepest near-IR observations \citep{illingworth13}. The ALMA observations were performed in both band 3 (3 mm) and 6 (1.2 mm), with the main aim of reconstructing the content of molecular gas and dust in galaxies over cosmic time through the detection of carbon monoxide (CO) lines and mm continuum emission \citep{decarli19,aravena20}.
 
The ASPECS 1.2 continuum observations reach an ultimate depth of $\sim$10 $\mu$Jy $rms$ \citep{aravena20}. We considered both the main sample of 35 high-fidelity ALMA detections and the secondary sample of 26 lower-fidelity ALMA detections with optical and/or near-IR priors, from which we removed galaxies without redshifts or with stellar masses below $10^{10}\;M_{\odot}$ (the upper bound of the sample is $M_*=3\times10^{11}\msun$). This leaves a total of 41 detected galaxies with redshifts (either spectroscopic or photometric) in the range $z=0.4-2.8$. The ISM masses $M_{gas}$ of ASPECS galaxies have been estimated in several ways, and the various methods produced results consistent with each other \citep{aravena20}. As described in Section~\ref{sect:gasmass}, we considered the $M_{gas}$ values derived from the dust masses assuming a fixed gas-to-dust mass ratio $\delta_{GDR}$. The selection function for the ISM (dust) mass of ASPECS galaxies is rather flat as a function of redshift (see e.g. Fig.~8 in \citealt{aravena20} and Fig~\ref{mdz} here): molecular gas masses down to $\sim 3\times 10^{9}\,M_{\odot}$ can be detected equally well at $z\sim 0.4-4$. The angular resolution of ALMA data in ASPECS is $\sim 1.3''$ FWHM, which is not enough to provide a reliable estimate of the extension of either the dust or molecular gas emission of high-$z$ galaxies. As an example, at the median redshift of the sample, $z=1.5$, ASPECS would only resolve galaxies with half-light radii $>5.5$~kpc, whereas the typical galaxy radius at that redshift is $\sim3$~kpc \citep{allen17}. We will then resort to galaxy sizes as estimated from stellar light in the optical-rest frame bands using the HST F160W photometry from CANDELS (\citealt{grogin11,koekemoer11}. We anticipate that a reliable optical size estimate is available for 32 galaxies out of the 41 massive galaxies in ASPECS (see Section~\ref{sect:galsize}).

\subsection{COSMOS $z\sim3.3$}\label{cosmos}

\citet{suzuki21} observed with ALMA a sample of 12 star-forming galaxies with spectroscopic redshift in the range $z_{spec}\sim 3.1-3.6$ in the COSMOS field \citep{scoville07}. Their targets were drawn from a parent sample of galaxies observed with near-IR spectroscopy, and were selected to have $M_*>10^{10}\;M_{\odot}$ and a reliable determination of their gas-phase metallicity.
The individual ALMA observations were performed at 1.3~mm (band 6) down to $10-50\;\mu$Jy $rms$ in the dust continuum, and 6 sources were detected at $>4\sigma$.
Similarly to the ASPECS sample, the angular resolution of the ALMA observations (average beam size $\sim1.4''$) is not sufficient to provide a reliable estimate of the source extent, 
and we will then resort to galaxy sizes as estimated from the available HST photometry (F814W band; \citealt{leauthaud07}).

\subsection{ALPINE}

The ALMA Large Program to INvestigate \cii\ at Early times (ALPINE; \citealt{lefevre20,bethermin20}) targeted 118 UV-selected normal star-forming galaxies with spectroscopic redshift in two windows, $z_{spec}\sim4.4-4.6$, and $z_{spec}\sim5.1-5.9$, in the COSMOS and ECDFS \citep{lehmer05} fields. The main aim of ALPINE is probing obscured star formation in distant galaxies by means their \cii\ line and dust continuum emission. The observations were performed in band 7 at a typical wavelength and down to an average continuum $rms$ of 850~$\mu$m and 28~$\mu$Jy for the $z\sim4.5$ targets,  and 1.05~mm and 50~$\mu$Jy for the $z\sim5.5$ targets, respectively \citep{bethermin20}. The \cii\ line was detected in 75 targets, and the continuum emission in 23 targets. Here we consider the 27 ALPINE galaxies with $M_*>10^{10}\;M_{\odot}$: 21 are detected in \cii, and 13 in the continuum (all of the continuum detected galaxies except one are also detected in \cii). The average beam size of the ALMA observations is $\sim0.9''$. This allowed \citet{fujimoto20} to measure the size of the \cii\ emission in the visibility plane, and derive a reliable size estimate for 9 of the massive galaxies considered here. In addition \citet{fujimoto20} derived optical size estimates for their targets based on HST F814W and F160W photometry: for example 7 out of the 13 ALPINE massive galaxies detected in the continuum feature a reliable HST size estimate (see Sect.~\ref{sect:galsize}). 
 
\subsection{Hosts of $z\sim6$ QSOs} 
 
At $z\sim 6$, the selection of massive galaxies with $M_*>10^{10-11}\msun$ is difficult because of their low space density \citep{grazian15,stefanon21}. This requires sampling large sky areas down to very faint magnitudes to allow the collection of sizable source samples. A possible alternative is using bright QSOs as beacons to reveal the position
of distant and massive galaxies to be followed up with dedicated ALMA observations. In fact, based on their large dust content and dynamical mass estimates, $z\sim 6$ QSOs are believed to be hosted by galaxies with stellar masses of least $\sim10^{10-11}\msun$ \citep{calura14, venemans18, izumi19}.
  
We considered the QSO sample published in \citet{venemans20}, who compiled a series of ALMA observations of 27 UV-bright ($M_{UV}\lesssim-25$) QSOs at $z_{spec}>5.7$ drawn from the major wide-area optical (SDSS, PanSTARRs, CFHQS) and near-IR (UKIDSS, VIKING) surveys. All targets were previously detected in both \cii\ and 1.2~mm continuum emission, and have $L_{\cii}\gtrsim10^9\lsun$.
The ALMA observations reported by \citet{venemans20} have a continuum sensitivity varying from $rms\sim$50 to $\sim$200 $\mu$Jy, and an average angular resolution of $\sim0.3''$ FWHM, significantly improving over past observations. Such sub-arcsec resolution allowed \citet{venemans20} to determine the spatial extent of both continuum and \cii\ emission. The sample overlaps significantly ($>50\%$) with that presented in \citet{decarli18} and \citet{venemans18}, in which the only constraint required for ALMA observations was target visibility, and it is then not biased towards high dust continuum or \cii\ emission. Nonetheless, the vast majority ($>85\%$) of objects in \citet{decarli18} and \citet{venemans18} were indeed detected by ALMA in both \cii\ and continuum emission. We therefore consider the ALMA measurements reported in \citet{venemans20} as representative of the population of massive high-$z$ galaxies hosting luminous QSOs. 
 
The Subaru High-$z$ Exploration of Low-Luminosity Quasars project (SHELLQs, \citealt{matsuoka16,matsuoka19}) is searching for faint QSO at $z\sim 6$ over the 1400 deg$^2$ Wide area of the Subaru Hyper Suprime-Cam (HSC) Strategic Program survey \citep{aihara18}. SHELLQs is currently providing the faintest samples of UV-selected QSOs at $z\sim6$, reaching absolute magnitudes of $M_{UV}\sim-22.8$. \citet{izumi18} and \citet{izumi19} followed up with ALMA a total of 7 SHELLQs QSOs with $-22.8<M_{UV}<-25.3$ in the redshift range $z_{spec}=5.9-6.4$. The observations were performed down to an average continuum $rms$ of $\sim$20~$\mu$Jy at 1.2~mm and with an average beam size of $\sim0.5''$. All QSOs were detected in the \cii\ line, and 6 out of 7 also in the continuum. In most cases the sub-arcsec angular resolution allowed a reliable determination of the source extent in both continuum and \cii\ (see Sect.~\ref{sect:galsize}). 

\begin{figure}[t]
\includegraphics[angle=0, width=9cm]{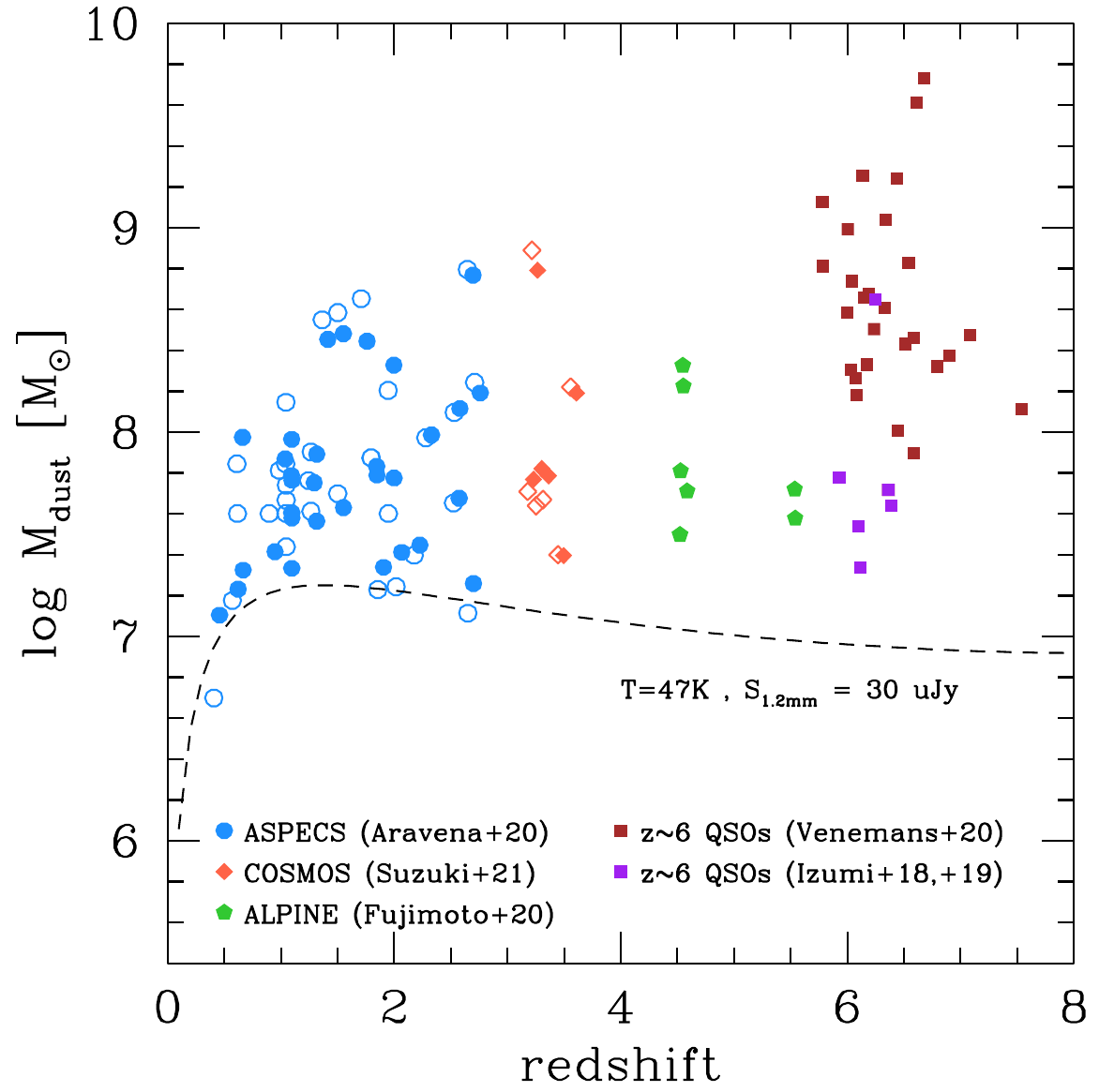}
\caption{Estimated dust mass vs redshift for the considered galaxy samples as labeled. For high-$z$ QSO hosts, dust masses have been estimated from Eq.~\ref{eq:mdust} assuming $T_d$=47~K if $T_{min}<47$~K or $T_d=1.2T_{min}$ if $T_{min}>47$~K (see text for details). For the ASPECS and COSMOS samples, the original dust masses derived through SED-fitting (\textsc{magphys}) are plotted as open symbols, and those derived from a single MBB with $T_d=47$~K as filled symbols. The dashed curve shows the dust mass limit corresponding to a flux density of 30 $\mu$Jy at 1.2~mm  (i.e. the $\sim3\sigma$ detection limit in ASPECS) from Eq.~\ref{eq:mdust} assuming $T_d$=47~K. 
} 
\label{mdz}
\end{figure} 

\begin{figure}[t]
\includegraphics[angle=0, width=9cm]{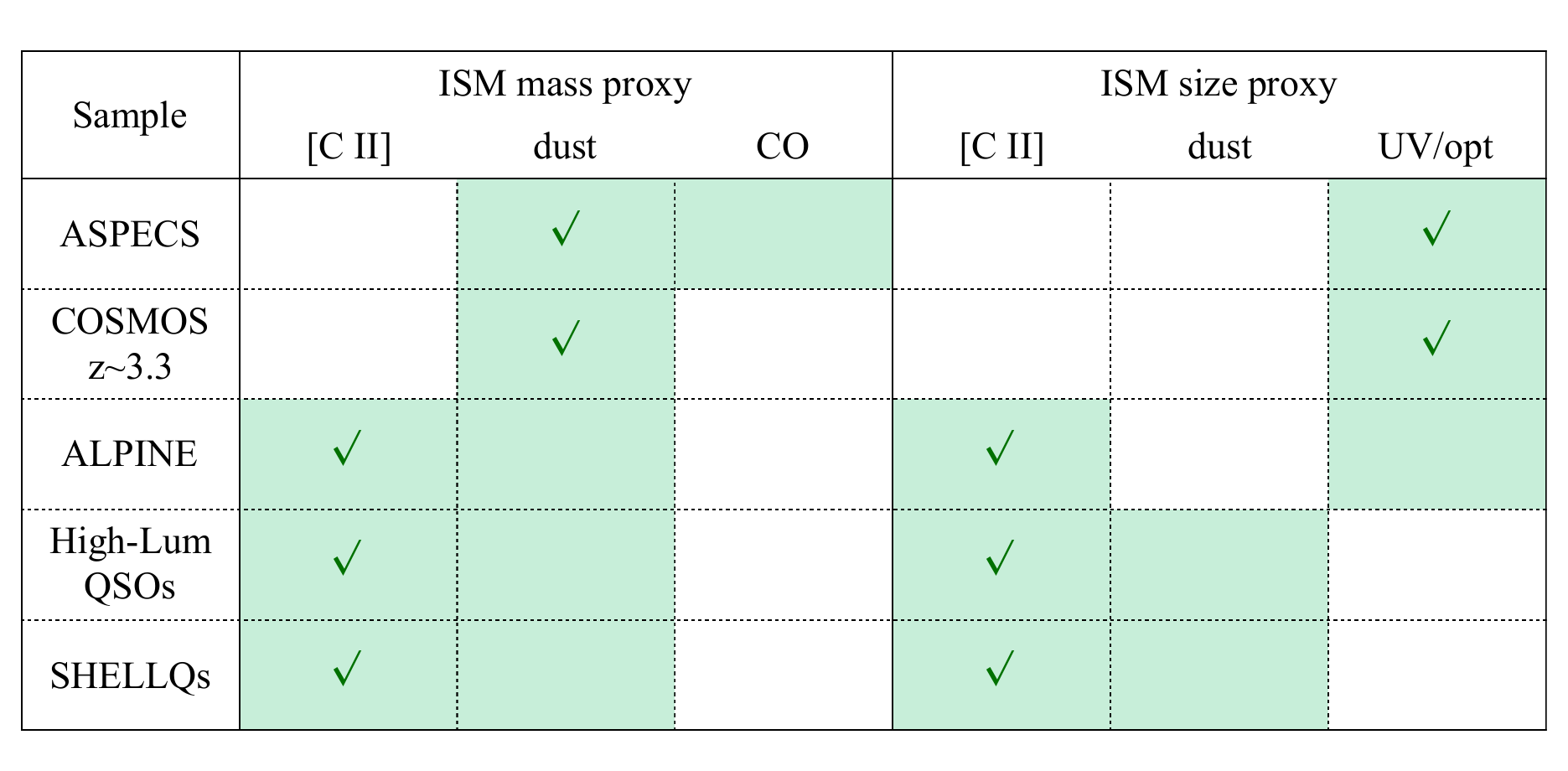}
\caption{Summary chart of the ISM mass and size proxies available for our samples (green shaded cells). Dark green ticks mark those proxies that have been used in the final ISM column density estimates (see Sect.~\ref{sect:gasmass} and \ref{sect:galsize} for details).
} 
\label{fig:green}
\end{figure} 

 \section{Methodology}\label{methods}
  
In this Section we describe the main data and calibrations used in the literature to derive ISM masses and sizes, i.e. the basic quantities needed to compute the ISM column density. We apply the different calibrations and cross-check their consistency on our samples. Fig.~\ref{fig:green} summarizes the ISM mass and size proxies that are available for the different samples. We anticipate that, whenever ALMA \cii\ data are available, these are at high spatial resolution. Therefore, we will use them to estimate both ISM masses and sizes. Otherwise, we will use dust continuum from ALMA to estimate ISM masses and optical/UV stellar emission from HST to estimate ISM sizes, as the spatial resolution of the available ALMA continuum data is often insufficient. This is detailed further below. 

\subsection {Gas mass estimates}\label{sect:gasmass}

The molecular gas content in galaxies can be estimated using different tracers. The most common are the dust continuum or CO emission lines corresponding to different rotational transitions CO($J\rightarrow J-1$), where $J$ is the rotational quantum number. ASPECS galaxies feature a high rate of CO detections and have well-sampled FIR SEDs, which allowed \citet{aravena20} to measure their gas masses $M_{gas}$ using three different methods: i) they converted the luminosity of different CO transitions $L'_{\rm CO(J\rightarrow J-1)}$ into the ground state line luminosity $L'_{\rm CO(1-0)}$ based on the average CO line ratios measured in their survey \citep{boogard20}, and then applied a standard calibration $M_{gas}=\alpha_{CO} L'_{\rm CO(1-0)}$ with $\alpha_{CO}=3.6$; ii) the dust masses derived from broad band SED-fitting with {\sc magphys} \citep{dacunha08,dacunha15} were converted into total gas masses as $M_{gas}=\delta_{GDR} M_d$ assuming a gas-to-dust mass ratio $\delta_{GDR}=200$, which is appropriate for objects with slightly sub-solar metallicity as those in their sample \citep{remy14,devis19}; iii) directly from the observed 1.2~mm flux density assuming the dust continuum luminosity - $M_{gas}$ calibration of \citet{scoville16}. The three methods found results consistent with each other within a factor of $\sim 2$ (see \citealt{aravena20}). 

Recently, a new empirical calibration based on the \cii\ line luminosity has been also proposed to track the molecular gas content of galaxies \citep{zanella18}. Because of the Carbon's low ionization potential (11.3 eV), C$^+$ may be present in different phases of the ISM, but most of the \cii\ emission is expected to emerge from the molecular phase (see \citealt{zanella18} and references therein). The \cii\ line, however, falls in the wavelength range where ALMA  observations are mostly efficient (band $\leq7$) only at $z\gtrsim4$, and hence no \cii\ data are available for ASPECS galaxies, nor for the COSMOS $z\sim 3.3$ sample. On the contrary, CO line observations are very sparse for our $z>4$ samples, and the few detections refer to lines from very high-$J$ rotational transitions, which are not reliable tracers of the molecular gas content. To compare ISM masses based on the same diagnostics, we then first investigate the dust content of galaxies in our samples, as continuum observations are available at all redshifts.

\citet{aravena20} and \citet{suzuki21} measured the dust masses of galaxies in the ASPECS and COSMOS $z\sim3.3$ sample, respectively, by means of broad band SED fitting with {\sc magphys}. 
For ALPINE and for high-$z$ QSO hosts, dust masses were obtained by normalizing a modified black body spectrum to single-band ALMA photometry, but using different dust temperatures: $T_d$=25~K for the ALPINE sample \citep{pozzi21} and $T_d$=47~K for high-$z$ QSO hosts \citep{venemans20,izumi18,izumi19} 
As the total estimated mass decreases with increasing dust temperature and there are no stringent constraints on the dust temperature for most of the considered galaxies, we
recomputed all dust masses following the procedure described below (see also \citealt{walter22}).

We considered a general modified black body (MBB) spectrum of the form:

\begin{equation}
S_{\nu_{obs}} = \Omega [B_{\nu}(T_d)-B_{\nu}(T_{CMB})] (1 - \mathrm{e}^{-\tau_{\nu}})(1+z)^{-3}, 
\label{snu}
\end{equation} 

where $\Omega$ is the solid angle covered by the source, $B_{\nu}\,(T) = \frac{2h\nu^{3}}{c^2}(\mathrm{e}^{\frac{h\nu}{k_B T}} - 1)^{-1}$ is the Planck function, $T_d$ is the dust temperature, $T_{CMB}(z)=2.725(1+z)\,\rm{K}$ is the CMB temperature at redshift $z$, and $\tau_{\nu}$ is the dust optical depth. In the above equation $\nu_{obs}$ and $\nu$ are the observed and rest frequency, respectively [$\nu=\nu_{obs}(1+z)$]. By considering that the dust optical depth is proportional to its surface density $\Sigma_d$, i.e. $\tau_{\nu}=\kappa_{\nu}\Sigma_d$, where $\kappa_{\nu}$ is the dust absorption coefficient, and that $\Sigma_d=M_d/A$, where $M_d$ is the dust mass and $A$ is the source physical area, once the FIR flux density, area and temperature of the dusty source are known, one can solve Eq.~\ref{snu} for $\tau_{\nu}$ and in turn determine the dust surface density and mass, which reads as:

\begin{equation}
M_d  = - {\rm ln}\left\{ 1 - \frac{ (1+z)^3 \,S_{\nu_{obs}} } {\Omega \, [B_{\nu}(T_d)-B_{\nu}(T_{CMB})]}  \right\} \, \frac{A}{\kappa_{\nu}} .
\label{eq:mdust}
\end{equation}

We note that, for $\tau_{\nu}<<1$ (and neglecting the correction for the CMB temperature), the above relation simplifies to the widely used formula \citep{hughes97,gilli14,venemans18,pozzi21}:

\begin{equation}
M_{\rm d} = \frac{D_{\rm L}^2 S_{\nu_{obs}} }{(1+z) \kappa_{\nu}B_{\nu}(T_{\rm d})},
\end{equation}

where $D_L$ is the luminosity distance (we recall that $D_L = (1+z)^2 D_A$, and $D_A^2=A/\Omega$ is the angular diameter distance).

As the argument within the curly brackets of Eq.~\ref{eq:mdust} needs to be positive, one gets:

\begin{equation}
B_{\nu}(T_d) > I_{\nu, \, min} \equiv \frac{(1+z)^3\,S_{\nu_{obs}}}{\Omega} + B_{\nu}(T_{CMB}) , 
\label{bcond}
\end{equation}

which in turn translates into a minimum temperature condition $T > T_{min}$, where:

\begin{equation}
T_{min} \equiv \frac{h \,\nu}{k_B}\frac{1}{{\rm ln}\left[ 1 + \frac{2h \nu^3}{c^2\,I_{\nu, \, min}} \right] }
\label{eq:tmin}
\end{equation}

Physically, this means that the larger the observed surface brightness $S_{\nu_{obs}}/\Omega$, the hotter the dust needed to explain it.
Eq~\ref{snu} shows that, for a given dust temperature, the observed surface brightness increases with increasing dust optical depth, i.e. when the source approaches an optically thick black body. In this case, only lower limits to the dust mass can be obtained.
However, Eq.~\ref{bcond} shows that dust of a given temperature cannot produce arbitrarily large surface brightnesses, even for arbitrarily large dust masses.

The dust temperature and emissivity index $\beta$ cannot be determined for most galaxies in our samples, as these have single-band mm detections only (except for ASPECS). We then followed the literature on high-$z$ QSOs and assumed $T_d=47$~K and $\beta=1.6$ \citep{calura14, venemans18, izumi18, venemans20}.
We also assumed $\kappa_{\nu}(\beta) = 0.77 \left (\nu/352{\rm GHz} \right)^{\beta}{\rm cm^2\,g^{-1}}$ \citep{dunne00}. 

\begin{figure}[t]
\vskip -0.4cm
\includegraphics[angle=0, width=9cm]{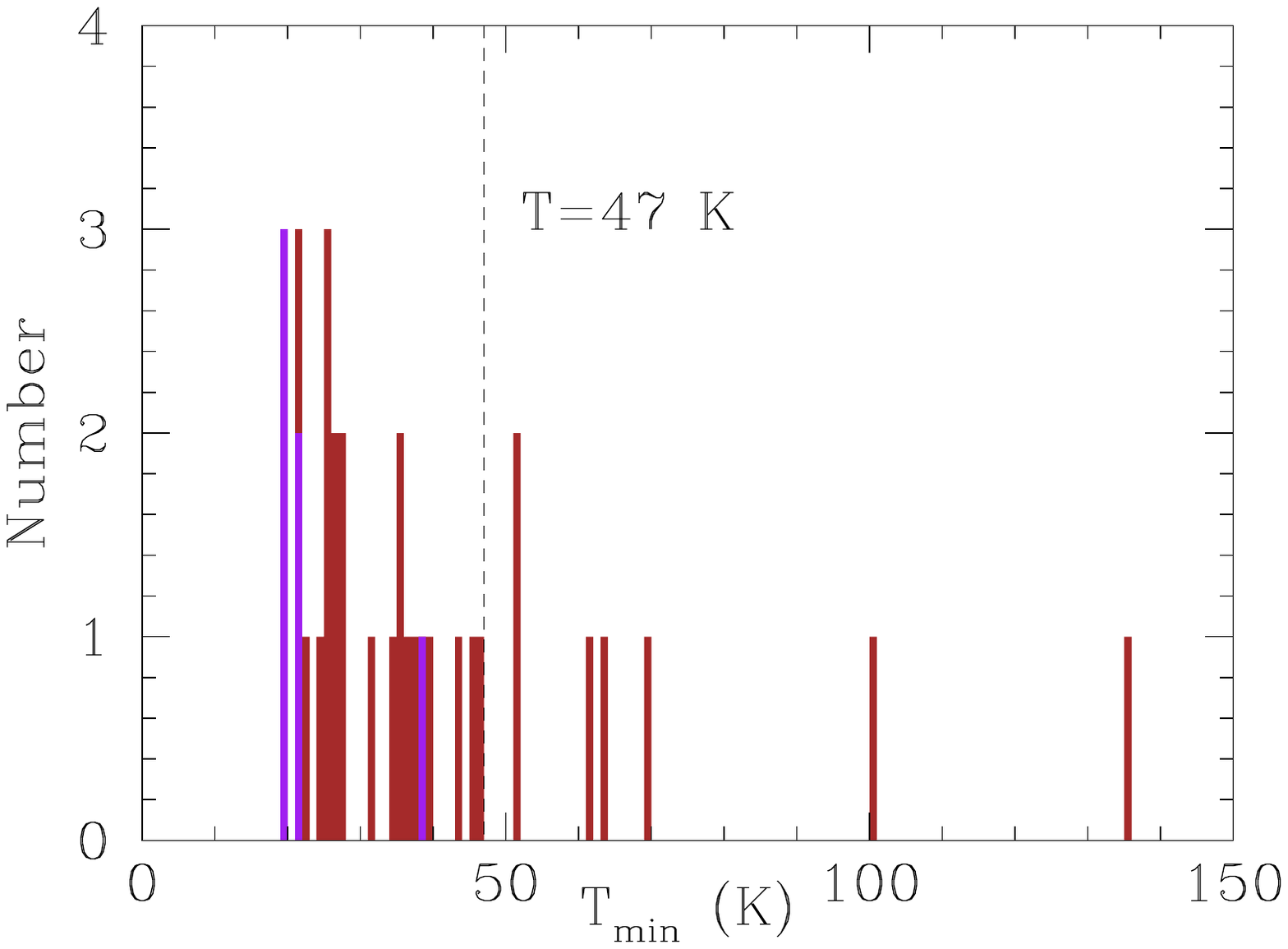}
\caption{Distribution of the minimum blackbody temperature $T_{min}$ required to explain the observed 1.2~mm brightness of $z\sim 6$ QSOs. The \citet{venemans20} and SHELLQs samples are shown in brown and purple, respectively. Note that the 1.2~mm emission on $\sim$kpc-scales of a few bright objects can only be reproduced by hot dust with $T_d>50-100$~K. We used $T_d=47$~K, a standard reference value used in the literature (e.g. \citealt{venemans20}), to derive the dust masses of all QSOs with $T_{min}<47$~K. For hotter systems, we used $T_d=1.2 T_{min}$ (see text for details).} 
\label{fig:tmin}
\end{figure} 

\begin{figure*}[t]
\includegraphics[angle=0, width=18.5cm]{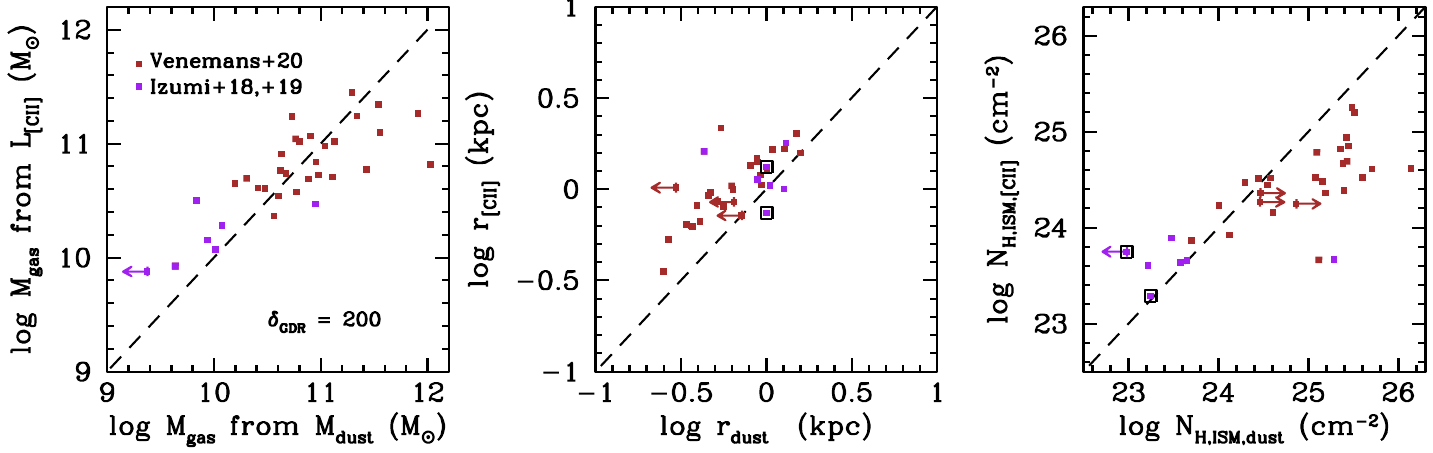}
\caption{Comparison between ISM mass ($left$), size ($center$) and column density ($right$) as derived from \cii\ and dust continuum observations for the $z\sim6$ QSO samples. The ISM masses derived with the two methods agree for $\delta_{GDR}=200$. The [C \textsc{ii}]-based sizes are on average $\sim 1.5$ times larger than the dust continuum sizes. The total ISM column densities are on average higher for the dust continuum method, owing to the smaller sizes measured. Three sources with upper limits to their continuum size in the \citet{venemans20} sample are marked with arrows in the $central$ and $right$ panels. One SHELLQs source is not detected in the continuum, and is shown with an arrow in the $left$ and $right$ panels. For this and for another source in the same sample (both marked with large open squares in the $central$ and $right$ panels), the size could not be estimated, so we assumed it equal to the median of the sample.} 
\label{cfr_z6qso}
\end{figure*} 

Fig.~\ref{fig:tmin} shows the distribution of the minimum blackbody temperature $T_{min}$ needed to explain the 1.2~mm surface brightness of high-$z$ QSOs obtained using 
Eq.~\ref{eq:tmin} and the dust continuum flux densities and sizes measured by \citet{venemans20,izumi18,izumi19}: the mm/FIR emission on $\sim$kpc-scales of a few bright and 
compact systems can be reproduced only by hot dust with $T_d>50$~K, or even $T_d>100$~K for the two most extreme cases, namely PSO231-20 and J2348-3054 in 
\citealt{venemans20}, which, in fact, feature the smallest sizes and larger surface brightnesses among all QSOs in that sample. This is supported by the results of \citet{walter22}, who 
recently observed the $z=6.9$ QSO J2348-3054 with ALMA in band 6 at 0.035'' (200pc) resolution, and derived its FIR SED by combining Herschel and lower resolution ALMA data in 
different bands. From the FIR SED, they measured a dust temperature of $T_d\sim 85$~K averaged over the whole system size of $\sim 1$~kpc, and through the high-resolution ALMA 
data they found that dust in the central regions is optically thick, and its minimum temperature increases to $T_{min}\sim130$~K within the inner $\sim$100 pc.
We then used $T_d=47$ K to derive the dust masses of all QSOs with $T_{min}<47$ K, and assumed $T_d=1.2 T_{min}$ for hotter systems. The latter choice ensures a good
agreement between the dust-based and the \cii-based gas mass estimates in these hot systems (see below). In fact, as the assumed temperature increases, the total dust mass -- and, in 
turn, the total gas mass -- decreases: dust-based gas masses become two times smaller than \cii-based masses already for $T_d>1.5 T_{min}$.

The dust masses derived for $z\sim6$ QSOs are shown in Fig.~\ref{mdz}. They span a very wide mass range from $\sim10^7$ to $\sim10^{10}\msun$, with bright QSOs having the larger dust contents. We applied the same method to compute dust masses for the other samples considered in this work, using the continuum flux densities as measured by ALMA and the source size derived from HST photometry (see Sect.~\ref{sect:galsize}),
and compared the results with those published in the literature. We found an excellent agreement between our estimates obtained with $T_d=47$~K and those obtained with {\sc magphys} for the ASPECS and COSMOS samples (see Fig.~\ref{mdz}). For the ALPINE sample, the dust masses published by \citet{pozzi21} using a modified black body with $T_d=25$~K are a factor of 4 higher than those estimated here. This is mainly related to the cold dust temperature used in \citet{pozzi21}: repeating the computation with $T_d=25$~K we  obtained dust masses in excellent agreement with those of \citet{pozzi21}. 
Fig.~\ref{mdz} also shows the $3\sigma$ detection limit in dust mass for the ASPECS sample ($rms\sim10\;\mu$Jy), which is $M_d\sim 2\times10^{7}\msun$ at $z=1$ and declines slowly towards higher redshifts. The typical detection limits of the other samples  are from $\sim 2$ (ALPINE, COSMOS, SHELLQs), to $\sim10$ (high-lum QSOs) times higher than the ASPECS limit.


We converted the dust masses into total gas masses by assuming $\delta_{GDR}=200$, as done by \citet{aravena20} for the ASPECS sample. Such a gas-to-dust mass ratio is in agreement with the expectations by chemical evolution models of evolved, massive disk galaxies with total baryonic mass $M_{tot}>10^{10}\msun$ \citep{calura17}, and with what is expected based on the metallicity of $M_*>10^{10}\msun$ galaxies at $z\sim 6$ in recent simulations \citep{torrey19}. 

In addition, we derived the total gas mass of ALPINE galaxies and of $z\sim6$ QSO hosts by means of the \cii\ line luminosity and the \citet{zanella18} calibration: $M_{gas}/\msun = 31 \times L_{\cii}/\lsun$. The [C \textsc{ii}]-based gas masses are generally consistent with those derived from the dust continuum (see left panel of Fig.~\ref{cfr_z6qso}).

\subsection {Galaxy size estimates}\label{sect:galsize}

The relation between the galaxy size as measured though the stellar light, dust emission, or gas density profiles is subject of intense study. The effective radii found
from HST photometry of distant galaxies (UV/optical rest-frame, i.e. tracking the stellar light) are on average $\sim$1.5-1.7 times larger than those estimated with ALMA for the dust continuum emission (e.g. \citealt{fujimoto17,lang19}), whereas they are similar to those derived for the cold gas distribution, e.g. from CO data \citep{calistro18}. The reasons for this discrepancy could be ascribed to dust temperature and optical depth gradients, which may cause a steeper drop in the FIR/mm continuum emission at large radii \citep{calistro18}. If so, stellar emission would be a more reliable tracer of the spatial distribution of cold molecular gas, i.e. of the bulk of the ISM, than the dust FIR-continuum emission. Recently, the effective radius of the \cii\ line-emitting gas was measured by \citet{fujimoto20} for the ALPINE sample. For $M_*>10^{10}\;M_{\odot}$ galaxies, the size of the \cii\ emitting region was found to be on average larger by a factor of $\sim3$ than the HST UV/optical-rest size. \cii\ emission traces both atomic and molecular gas, and atomic gas is known to be more diffuse (at least in local galaxies; \citealt{bigiel12}).
Furthermore, large-scale, star-formation driven outflows of neutral gas are often observed in massive galaxies both in the local \citep{cazzoli16,roberts20} and distant \citep{ginolfi20}  Universe, which may also explain the larger \cii\ sizes.

The resolution of the ALMA data available for both ASPECS and COSMOS $z\sim3.3$ galaxies ($1.3-1.4''$ FWHM) is not sharp enough to provide a reliable estimate of their cold gas extension. We will then rely on UV/optical rest-frame measurements based on HST data. 
For ASPECS, we cross correlated the \citet{aravena20} sample with that presented by \citet{vanderwel12}, who run GALFIT \citep{peng02,peng10} to measure the structural parameters of galaxies in the CANDELS fields selected by means of HST F160W photometry. 
The structural parameters relevant for this study are the effective radii $r_e$ (actually, the half-light semi-major axes), Sersic indices $n$, and axial ratios $q$. The distributions of Sersic indices of the whole sample of \citet{vanderwel12} and of the ASPECS subsample both peak at $n=1$. 
We considered only the 32 ASPECS galaxies (23 from the main sample, and 9 from the secondary sample) where GALFIT returned a good fit (quality flag = 0). Typical sizes are $r_e\sim3-4$~kpc. For the COSMOS $z\sim3.3$ galaxies, we cross-correlated the \citet{suzuki21} sample with the COSMOS HST/ACS galaxy catalog \citep{leauthaud07}, which contains $1.2$ millions galaxies detected in the $z_{814}$ band down to 26.5 AB mag, and reports information about their half-light radii as measured from HST data. The UV sizes of galaxies in the \citet{suzuki21} sample are in the range $r_e\sim1-2.5$ kpc.

For the ALPINE sample (average ALMA beam size $\sim 0.9''$), there are no public estimates of the size of the dust continuum emission, whereas \cii\ sizes are available \citep{fujimoto20} and will be used in the following. We also note that \citet{fujimoto20} derived optical/UV sizes for their targets using GALFIT on HST F814W and F160W photometric data. In particular, they were able to obtain a reliable size measurement (flag=0) for 7 massive ALPINE galaxies detected in the continuum (see Fig.~\ref{mdz} and Table~\ref{tab:sample}).

For the sample of high-$z$ QSO hosts, ALMA observations have been performed with a resolution of 0.3-0.5'', which is sufficient to resolve both dust and \cii\ emission in most cases. The comparison between the [C \textsc{ii}]-based and dust-based effective radii is shown in Fig.~\ref{cfr_z6qso} ($center$). On average, 
the [C \textsc{ii}]-based radii are $\sim1.5$ larger than dust-based radii. We recall that, in known high-$z$ QSOs, the nuclear radiation largely outshines that of stars at optical/UV rest-wavelengths. As a result, in spite of the few attempts to separate the QSO light from that of its host through high-resolution HST observations \citep{mechtley12,marshall20}, no information is currently available about the stellar emission of $z\sim6$ QSO hosts.

\section{ISM absorption estimates}

\subsection{ISM density profile and column towards galaxies' nuclei}\label{elmet}

To estimate the ISM column density towards galaxy nuclei, we approximate galaxies as disks where both the light surface brightness and the gas density profiles decline exponentially along the radial direction. This is equivalent to assume a Sersic index of $n=1$, which is consistent with the average light surface brightness profiles measured in the ASPECS and CANDELS samples. Furthermore, recent ALMA observations at 0''.07 resolution of a small sample of distant, sub-millimeter galaxies \citep{hodge19} showed that their continuum emission is dominated by exponential dust disks, which also supports our assumptions. Given an exponential profile of the form:

\beq
I(r) =  I_0 \; e^{-r/r_0}, 
\label{surbri}
\eeq

where $I_0$ is the central surface brightness and $r_0$ is the scale ($e$-folding) radius, it is convenient to rewrite Eq.~\ref{surbri} in terms of the effective, or half-light radius $r_e$, defined as the radius enclosing half of the total luminosity:

\beq
\int_0^\infty I(r)2\pi r \,dr = 2 \int_0^{r_e} I(r)2\pi r \,dr \;.
\label{re}
\eeq

Eq.~\ref{surbri} then reads as:

\beq
I(r) =  I(r_e) \; e^{-(r-r_e)/r_0} \; .
\label{surbrire}
\eeq

\begin{figure}[t]
\includegraphics[angle=0, width=9cm]{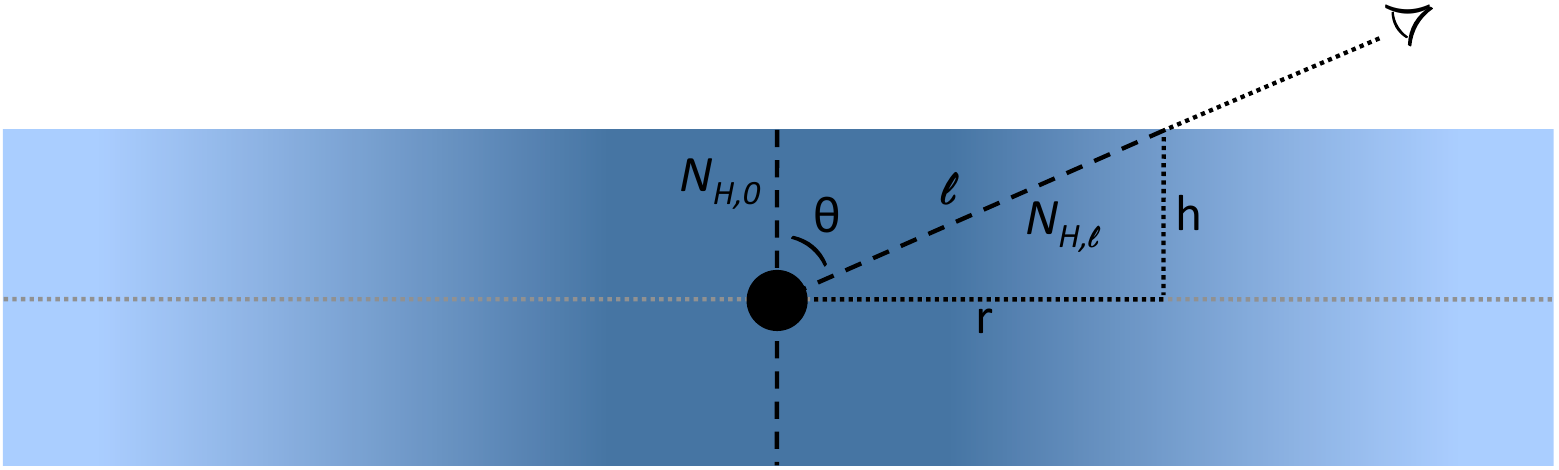}
\caption{Edge-on sketch of the adopted disk galaxy geometry. The darker shading reflects the higher ISM density of the assumed exponential profile.} 
\label{fig:scheme_smooth}
\end{figure}

For a pure exponential disk, $r_e=1.678 \, r_0$.
We then consider a simple disk model for the gas distribution and work in cylindric coordinates. In the radial direction, we assume that the gas follows the same exponentially decreasing profile as the light surface brightness. For simplicity, we assume that there is no density gradient along the vertical ($z$) direction. A sketch of the adopted geometry is shown in Fig.~\ref{fig:scheme_smooth}. Denoting $2h$ as the disk thickness, the total gas surface density is:

\beq
\Sigma_{gas}(r) =  \int_{-h}^{h} \rho(r,z)dz = 2h \rho_0 e^{-r/r_0} = \Sigma_{gas,0}\,e^{-r/r_0}, 
\label{surden}
\eeq

where $\rho_0$ is the central gas volume density and $\Sigma_{gas,0}=2h \rho_0 $. By rewriting Eq.~\ref{surden} in terms of the effective radius, one gets:

\beq
\Sigma_{gas}(r) = \Sigma_{gas}(r_e)\,e^{-(r-r_e)/r_0} \; .
\label{surdenre}
\eeq

Observationally, from the total gas mass $M_{gas}$ and effective radius $r_e$ measured for a given galaxy, one can derive an average value for the gas surface density as:

\beq
\Sigma_{gas} = (M_{gas}/2)/(\pi r_e^2) \; .
\label{avo}
\eeq

This quantity can be easily related to the central gas surface density  $\Sigma_{gas,0}$ by considering that, by definition:

\beq
M_{gas} =  \int_0^{\infty} \Sigma_{gas}(r)2\pi r\,dr  \; .
\label{mg}
\eeq

By combining Eqs.~\ref{surden}, \ref{avo}, and \ref{mg}, one derives: 

\beq
 \Sigma_{gas,0} =  (r_e/r_0)^2\, \Sigma_{gas} \sim 2.8\,\Sigma_{gas} \; .
\label{avrel}
\eeq

Then, for a face-on view of a pure exponential disk galaxy one would measure a column density towards the nucleus equal to:

\beq
 N_{H,0} = \Sigma_{gas,0}/2  \sim 1.4 \, \Sigma_{gas}  \; ,
\label{nzero}
\eeq

whereas for a random orientation angle $\theta$ between the galaxy axis and the line of sight, one has to integrate the density profile along the optical path $\ell$ (see the sketch in Fig.~\ref{fig:scheme_smooth}):

\beq
N_{H,\ell} =  \int_0^{\ell} \rho(r,z)d\ell = \int_0^{h\,tan\theta} \rho_0 e^{-r/r_0}\frac{dr}{sin\theta} \; .
\label{nhl}
\eeq

This gives:

\beq
N_{H,\ell} = \frac{r_0 \rho_0}{sin\theta} [ 1 - e^{-\frac{h}{r_0}tan\theta}]  \; , 
\label{nhangle}
\eeq

which increases from $h \rho_0$ to $r_0 \rho_0$ going from a face-on ($\theta=0^o$) to an edge-on ($\theta=90^o$) view. 
For random disk orientations, the average viewing angle is 1 radian, i.e. $\sim 57.3^{\rm o}$. By assuming a characteristic disk half scale-height $h=0.15 r_e$ \citep{elmegreen06,bouche15} and $\theta=57.3^{\rm o}$, from Eq.~\ref{nhangle} one then finds a column density towards the nucleus equal to: 

\beq
N_{\rm H,ISM}\sim1.5 \,N_{H,0} \sim 2.1 \, \Sigma_{gas}  \; .
\label{nhave}
\eeq

Therefore, for a purely exponential disk with random orientation, the average ISM column density towards the nucleus \nhism\ can be estimated as about twice the measured mean gas surface density $\Sigma_{gas}$. 

\begin{figure}[t]
\includegraphics[angle=0, width=9cm]{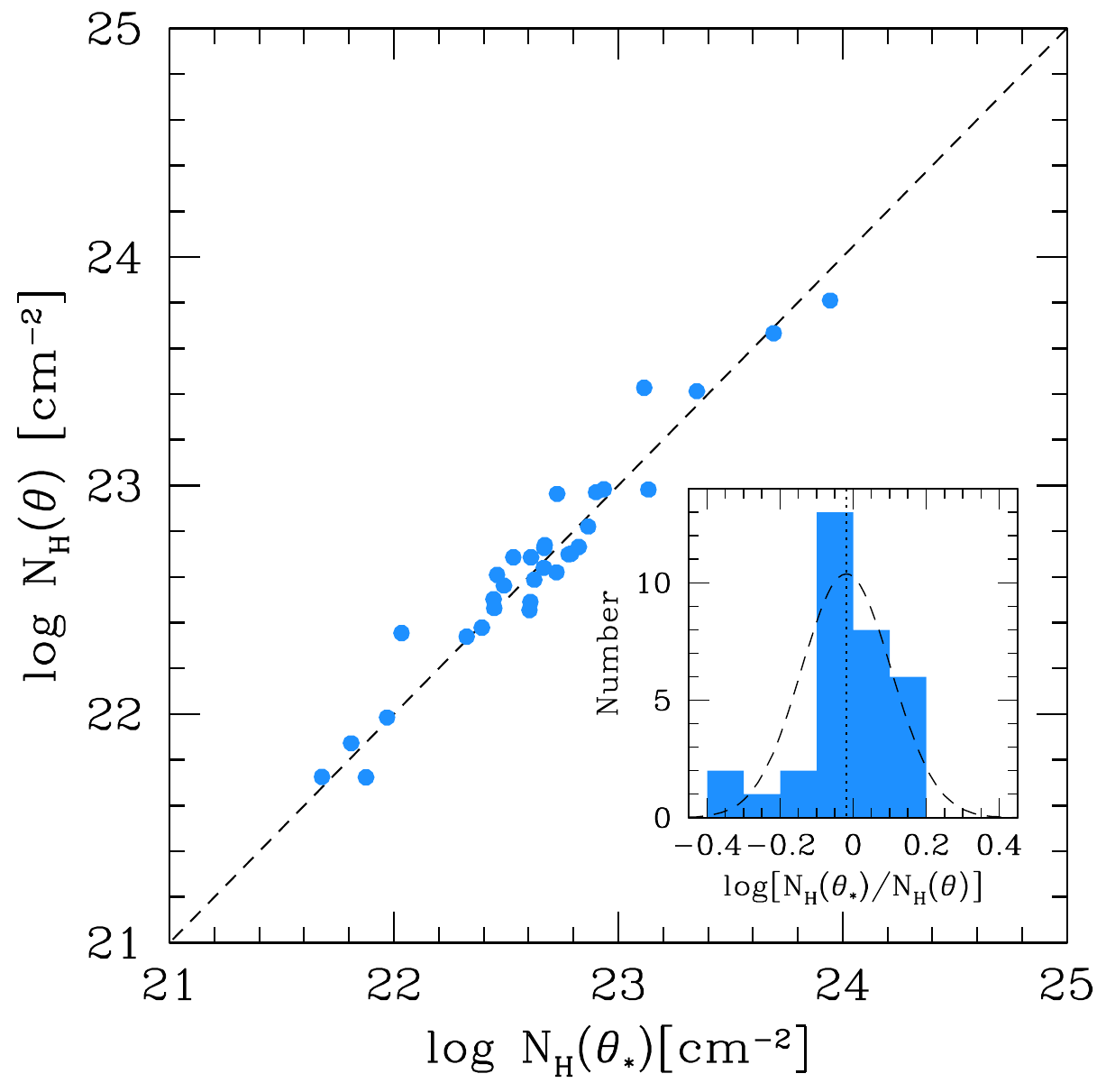}
\caption{Comparison between the ISM column densities towards the nucleus of ASPECS galaxies derived by assuming a fixed, average galaxy orientation ($\theta_*$=57.3 deg) vs those derived from individual galaxy orientations $\theta$. The inset shows the distribution of the $\Delta$log\nhism\ values: a Gaussian fit (dashed curve) is centered at about 0 (dotted line) and has $rms \sim 0.15$ dex.} 
\label{nhangle_all}
\end{figure} 

We computed the column density towards the nucleus of ASPECS galaxies based on both Eq.~\ref{nhangle} and Eq.~\ref{nhave}. In the thin-disk approximation, the orientation angle of each galaxy can be derived as $\theta = arcos(q)$, where $q$ is the axial ratio reported by \citet{vanderwel12} based on CANDELS F160W photometry. We note that the average axial ratio measured for ASPECS galaxies would correspond to $\theta\sim 55^{\rm o}$, i.e. remarkably similar to that expected for randomly oriented thin disks. The comparison between the column density obtained using a fixed, average orientation (Eq.~\ref{nhave}) vs those obtained from the individual orientations (Eq.~\ref{nhangle}) is shown in Fig.~\ref{nhangle_all}. The two estimates are generally within a factor of $\sim2$, and the $\Delta$log\nhism\ distribution has mean around 0 and $rms\sim0.15$ dex. We therefore deem using a fixed, angle-average orientation a sufficiently good approximation for our purposes and utilize Eq.~\ref{nhave} to derive the ISM column densities in all the considered galaxy samples. 

In Fig.~\ref{cfr_z6qso} ($right$) we show the comparison between the high-$z$ QSO ISM column densities estimated using the [C \textsc{ii}]-based ISM masses and sizes ($N_{H,ISM,\cii}$) vs those estimated with the dust-based gas masses and sizes ($N_{H,ISM,dust}$). The column densities obtained with the two methods are in reasonable agreement with each other, but a non-linear trend is visible, which is likely related to the overall system luminosity, as the high- and low-column regimes are mostly populated by high- \citep{venemans20} and low- (SHELLQs) luminosity QSOs, respectively. Indeed, because of the well known \cii\ luminosity deficit in FIR-bright systems \citep{malhotra01,graciacarpio11,diazsantos17,decarli18}, the ratio between the [C \textsc{ii}]-based and the dust-based ISM masses decreases towards high QSO luminosities. More powerful QSOs have also stronger \cii outflows \citep{bischetti19} that may enhance the size difference between the \cii\ and the dust emitting regions. Therefore, as the QSO luminosity increases, using the dust-based method one would derive larger ISM masses and smaller sizes than using the [C \textsc{ii}]-based method, obtaining in turn the observed trend between the two column density estimates.

\subsection{ISM clumpiness and obscured AGN fraction}\label{sub:clump}
 
In the previous sections we derived column density values assuming that the cold ISM is smoothly distributed across galaxy volumes. This is obviously not true: the molecular ISM phase, the one which is most concentrated and expected to produce most obscuration, is in fact distributed within dense clouds, which cover only a portion of the solid angle towards the nucleus. Here we aim to estimate the covering factor of these clouds and, in turn, the fraction of ISM-obscured AGN.

In the Milky Way, the average mass, radius and surface density of molecular clouds are $M_c=10^5\msun$, $R_c$=30 pc, and $\Sigma_c=28$\msunpc, respectively \citep{miville17}. Their volume filling factor $\phi=n_cV_c$ , where $n_c$ is the number of clouds per unit volume, and $V_c$ is the volume of each cloud, is about 1-2\%.
Such values may be radically different for massive galaxies at high redshift. As an example, the surface density of the individual molecular clouds detected in the ``Snake'' lensed galaxy at $z=1$ \citep{mirka19} is about 30 times higher than that of the Milky Way clouds.  Furthermore, the cloud size and filling factor may be larger, 
hence producing significant coverage of the nucleus. 

Similarly to what we did in the previous Section for a smooth gas distribution, we assume that the cloud number density is exponentially decreasing along the galaxy radius and is constant in the vertical direction, that is:
 
\beq
n_c(r,z) =  n_0 \, e^{-r/r_0} \; , 
\label{ncr}
\eeq

where $n_0$ is the number of clouds per unit volume at galaxy center.  
Following \citet{nenkova08}, the average number of clouds along a path $\ell$ at a given angle $\theta$ is given by:

\beq
\mathcal{N}_T(\theta) = \int N_c(\ell)d\ell \; ,
\label{nth}
\eeq

where $N_c=n_c A_c$ is the number of clouds per unit length and $A_c=\pi R_c^2$ is the cloud cross-sectional area. By integrating Eq.~\ref{nth} we derive:

\beq
\mathcal{N}_T(\theta) = \frac{\tau}{sin\theta} [ 1 - e^{-\epsilon\,tan\theta}]  \; , 
\label{nhcl}
\eeq

where $\tau=n_0 A_c r_0$ and $\epsilon=h/r_0$ (see the analogous Eq.~\ref{nhangle} for the total column density of a smooth medium). 

\begin{figure}[t]
\includegraphics[angle=0, width=8cm]{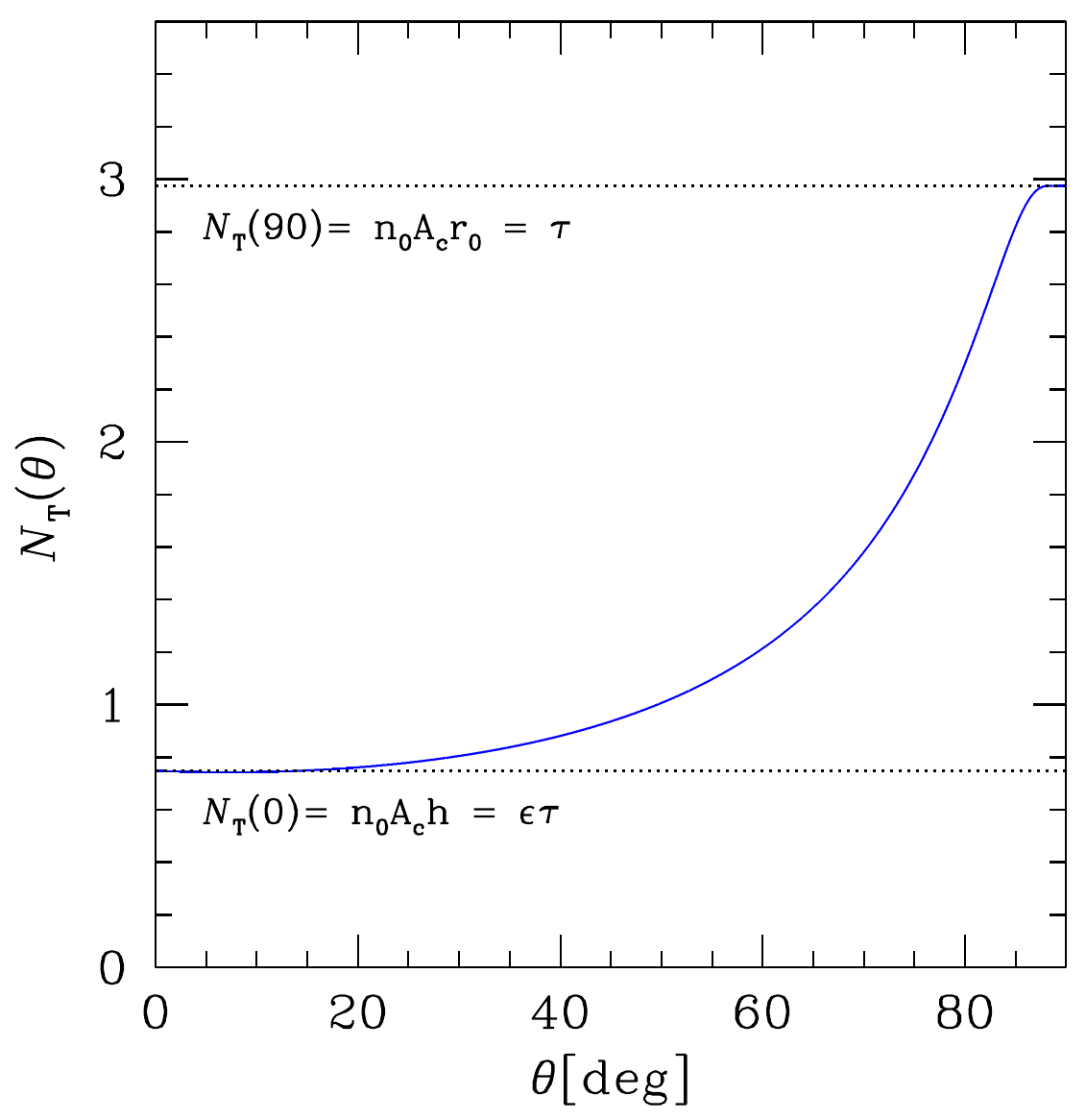}
\caption{Variation of the average number of ISM clouds along a given line of sight to the nucleus  as a function of the viewing angle $\theta$ ($90^o$ is edge on) for a typical massive galaxy with $M_{gas}\sim10^{10}\msun$ at $z=2.5$. } 
\label{nttheta}
\end{figure}

Based on Eq.~\ref{nhcl}, the average number of clouds seen face-on ($\theta=0$) or edge on ($\theta=\pi/2$) is $\mathcal{N}_T(0)=n_0 A_c h$ and $\mathcal{N}_T(\pi/2)=n_0 A_c r_0$, respectively.
So, the two parameters $\tau$ and $\epsilon$ in Eq.~\ref{nhcl} correspond to the average number of clouds seen edge-on and to the ratio between the average number of clouds seen face-on vs edge-on, respectively. The variation of $\mathcal{N}_T(\theta)$ with $\theta$ is shown in Fig.~\ref{nttheta} for a system with $n_0=270$ kpc$^{-3}$, $R_c=(A_c/\pi)^{0.5}=50$ pc , $\epsilon=0.25$, $r_e=1.678r_0=2.3$ kpc. These numbers are representative of high-$z$ galaxies. The assumed galaxy half-light radius $r_e$ is, in fact, typical of massive systems at $z=2.5$ \citep{allen17}. In addition, for $M_c=5\; 10^6\msun$ (which is a reasonable cloud mass at high-$z$, see the next Sections), the total gas mass in the system is $M_{gas}\sim10^{10}\msun$ (see e.g. Eq.~\ref{nontot}), which is again typical of massive galaxies at $z=2.5$ \citep{aravena20}. Clearly, the typical number of clouds intercepted along a given line of sight (e.g. Fig.~\ref{nttheta}) is subject to a series of uncertainties that increase when moving to progressively higher redshifts, where our knowledge of the average galaxy morphology (e.g. $\epsilon$) and of the individual ISM  cloud parameters (e.g. $R_c$) is poorer. Such uncertainties need to be kept in mind in the following analysis.

Once the average number of clouds along each line of sight is known, one can compute the total cloud covering factor to the nucleus, or, equivalently, the fraction of obscured nuclei $f_{obsc}$ in an unbiased sample of AGN \citep{nenkova08}:

\beq
f_{obsc}=1-\int_{0}^{\pi / 2} e^{-\mathcal{N}_{T}(\theta)} \sin \theta d \theta \; .
\label{fobsc}
\eeq  
  
\begin{figure}[t]
\includegraphics[angle=0, width=9cm]{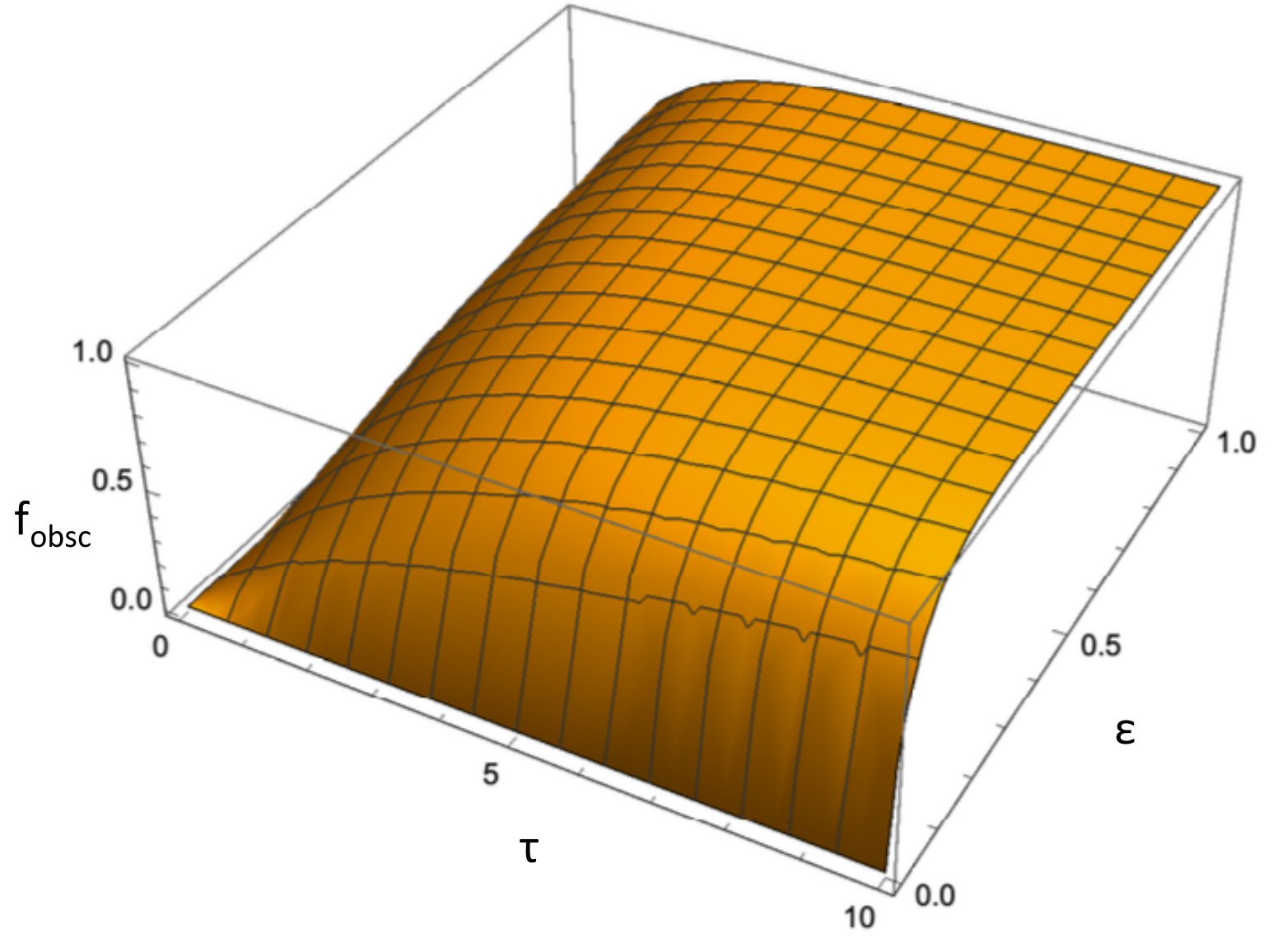}
\caption{Obscured AGN fraction as a function of $\tau$, the average number of clouds along edge-on views, and $\epsilon$, the ratio between the average number of clouds seen face-on vs edge-on .} 
\label{colin}
\end{figure}

We note that Eq.~\ref{fobsc} assumes that individual clouds are optically thick. At soft X-ray energies, this condition is in fact satisfied given their typical column densities. As an example, the typical column density of molecular clouds in the Milky Way is \nh$\sim 3\times10^{21}$\cm \citep{miville17}, meaning that they are optically thick to X-ray photons with $E<1$~keV. 
The dependence of $f_{obsc}$ on the two parameters $\tau$ and $\epsilon$ is shown in Fig.~\ref{colin}.

The cloud number density at the galaxy center $n_0$ can be related to the total number of clouds in the galaxy $N_{tot}$, and in turn to the total gas mass  $M_{gas}=N_{tot}M_c$, as follows:

\beq
N_{tot} = \int dz \int n_c(r,z)2\pi r dr  \; ,
\label{ntotint}
\eeq

which gives:

\beq
n_0 = \frac{N_{tot}}{4 \pi h r_0^2} = \frac{1}{4 \pi \epsilon r_0^3} \frac{M_{gas}}{M_c} \; .
\label{nontot}
\eeq

By considering that $\Sigma_{gas}=(M_{gas}/2)/(\pi r_e^{2})$, $r_e=1.678r_0$,  and $\Sigma_c  = M_c/A_c$, Eq.~\ref{nhcl}  can be written as:

\beq
\mathcal{N}_T(\theta) \sim 1.4 \frac{\Sigma_{gas}}{\Sigma_c}\frac{ 1 - e^{- \epsilon\, tan\theta} } {\epsilon\, sin\theta} \; .
\label{nhclgas}
\eeq

The main quantities that affect the average number of clouds observed in a given direction therefore depend only on the total galaxy gas surface density $\Sigma_{gas}$ (which is an observable, see previous Section), the gas surface density of each cloud $\Sigma_c$, and the galaxy ``aspect ratio'' $\epsilon$.
The meaning of Eq.~\ref{nhclgas} is clear: the number of clouds along a given direction increases as $\Sigma_{gas}$ increases, as clouds are closer in space. Also,
at a given $\Sigma_{gas}$, more clouds are intercepted along the line of sight if they are more diffuse, i.e. if $\Sigma_c$ decreases.

\subsection{ISM clouds with a distribution of physical parameters}\label{sub:clump2}

We now generalize the computations presented in Section~\ref{sub:clump} by moving from a single cloud type to a distribution of clouds with different radii $R_c$, masses $M_c$, and surface densities $\Sigma_c$. We assume that the shapes of these distributions are the same all over the galaxy volumes, and that only their normalizations, i.e. the cloud number density, varies following the radial exponential decline described in the previous Sections (see e.g. Eq.~\ref{ncr}).

We first considered the distributions observed for molecular clouds in the Milky Way by \citet{miville17}, who did not find any correlation between the cloud surface density and size. We then consider $\Sigma_c$ and $R_c$ as independent variables and write the number density of clouds at the galaxy center per unit cloud surface density and radius as:

\beq
n_o(\Sigma_c,R_c) = k\, n(R_c)n(\Sigma_c) \; ,
\label{nosr}
\eeq

where $n(R_c)$ and $n(\Sigma_c)$ are functions of only $R_c$ and $\Sigma_c$, respectively, and $k$ is a normalization constant.

By marginalizing over $\Sigma_c$ and $R_c$ one can write:

\beq
n_o(\Sigma_c)=k I(R_{c,*}) n(\Sigma_c) \; ,
\eeq

and

\beq
n_o(R_c)=k I(\Sigma_{c,*}) n(R_c) \; ,
\eeq

where $n_o(\Sigma_c)$ and $n_o(R_c)$ are the volume density of clouds at the galaxy center per unit $\Sigma_c$ and $R_c$, respectively, $I(R_{c,*}) =\int n(R_c)dR_c$, and $I(\Sigma_{c,*})=\int n(\Sigma_c)d\Sigma_c$.

\citet{miville17} divide clouds in the Milky Way into inner clouds and outer clouds, depending on whether their distance to the Galactic center is smaller or larger than 8.5 kpc, respectively.
The size distributions of inner and outer clouds both peak at $R_c\sim30$~pc and have similar shapes, whereas inner clouds are on average significantly denser and more massive, and contain $\sim85\%$ of the total gas mass in the \citet{miville17} catalog.
We then derive the shape of $n_o(\Sigma_c)$ and $n_o(R_c)$ , and hence of $n(R_c)$ and $n(\Sigma_c)$ by considering the $R_c$ and $\Sigma_c$ distributions of the inner Milky Way clouds, as these may be a better proxy of the clouds living in the compact environments of high-$z$ galaxies. We downloaded the cloud catalog presented by \citet{miville17}, and found that the $R_c$ and $\Sigma_c$ distributions can be well approximated by cut-off power-laws (Schechter functions) of the form:

\beq
n(R_c) = R_c^\alpha e^{-R_c/R_{c,*}} \; ,
\eeq

and 

\beq
n(\Sigma_c) = \Sigma_c^\alpha e^{-\Sigma_c/\Sigma_{c,*}} \; ,
\eeq

with $\alpha=1.7$, $R_{c,*}=18$~pc, $\beta=1.8$, $\Sigma_{c,*}=25$\msunpc.
With such approximations, $I(R_{c,*})$ and $I(\Sigma_{c,*})$ can be written as:

\beq
I(R_{c,*})=R_{c,*}^{\alpha+1}\Gamma(\alpha+1) \; ,
\eeq

and

\beq
I(\Sigma_{c,*})=\Sigma_{c,*}^{\beta+1}\Gamma(\beta+1) \; ,
\eeq

where $\Gamma(\xi)=\int_0^\infty x^{\xi-1}e^{-x}dx$ is the usual Gamma function.
The total cloud number density at the galaxy center can then be written as:

\begin{align}
n_0 = \int\int n_o(\Sigma_c,R_c)d\Sigma_c dR_c = k\,I(\Sigma_{c,*})\,I(R_{c,*}) = \\ \notag
k\,R_{c,*}^{\alpha+1}\Sigma_{c,*}^{\beta+1}\Gamma(\alpha+1)\Gamma(\beta+1) \; .
\end{align}

Similarly, by recalling that the mass of each cloud is $M_c=\Sigma_c \pi R_c^2$, the total gas mass density at the galaxy center can be written as:

\beq
\rho_0 = \int\int n_o(\Sigma_c,R_c) \pi \Sigma_c R_c^2 d\Sigma_c dR_c = \pi k\,R_{c,*}^{\alpha+3}\Sigma_{c,*}^{\beta+2}\Gamma(\alpha+3)\Gamma(\beta+2) \; .
\label{rhoo}
\eeq

By integrating over the galaxy volume (see Eqs. \ref{ntotint}, \ref{nontot}), the total number of clouds and gas mass in a galaxy are then:

\beq
N_{tot} = n_o 4 \pi h r_o^2 \; ,
\eeq

and

\beq
M_{gas} = \rho_o 4 \pi h r_o^2 \; .
\label{totmgas}
\eeq 

This simplified approach provides a satisfactory description of the overall cloud distribution and gas content of the Milky Way. As an example, 
for an input total gas mass equal to that contained in the inner Milky Way clouds ($\sim1.4\times 10^9\msun$, \citealt{miville17}), the model recovers
the total number of clouds within a factor of $\sim1.5-2$. Furthermore, the model reproduces the inner Milky Way cloud distribution in the $R_c$ vs
$M_c$ plane. The observed distribution is indeed well fit by the relation $M_c\propto R_c^{2.2\pm0.2}$ and peaks at [$R_c\sim35$pc, $M_c\sim3\times10^5\msun$],
whereas in the model $M_c\propto R_c^2$ by definition, and the distribution peak is found at [$R_c\sim30$pc, $M_c\sim1.5\times10^5\msun$].


\subsection{ISM-obscured AGN fraction for different \nh\ thresholds}\label{sub:clump3}

The formalism developed in the previous Sections allows us to compute the solid angle to galaxy nuclei intercepted by clouds above any given surface density threshold, or, equivalently, above any column density threshold $N_{\rm H,th}$. We can then compare our expectations with the fraction of AGN  with \nh~$>N_{\rm H,th}$ measured at different redshifts in wide and deep X-ray surveys. In addition, we can make forecasts about the obscured AGN populations at $z>5-6$ which are still missing from our cosmic inventory.

For a cloud population with a distribution of sizes and surface densities, the number of clouds with \nh~$>N_{\rm H,th}$ ($\Sigma_c>\Sigma_{th}$) per unit length can be written as:

\beq
N_{c,th} = \int_{\Sigma_{th}}^{\infty} \int_0^{\infty} n_c(\Sigma_c,R_c,r,z) \pi R_c^2 dR_c d\Sigma_c \; ,
\eeq

where $n_c(\Sigma_c,R_c,r,z)=n_o(\Sigma_c,R_c)e^{-r/r_o}$ is the number density of clouds per unit cloud radius and surface density at a given position in the galaxy $(r,z)$, and $n_o(\Sigma_c,R_c)$ is the number density of clouds per unit cloud radius and surface density at the galaxy center, as defined in Eq.~\ref{nosr}.

The average number of clouds with \nh~$>N_{\rm H,th}$ along a path $\ell$ at a given angle $\theta$ is then given by (see Eq.~\ref{nth}):

\begin{align}
\mathcal{N}_{T,th}(\theta) = \int N_{c,th}(\ell)d\ell = \\ \notag
\int_0^{\infty}d\ell \int_{\Sigma_{th}}^{\infty} \int_0^{\infty}  n_o(\Sigma_c,R_c)e^{-r/r_o}\pi R_c^2 dR_c d\Sigma_c \; ,
\end{align}

which leads to (see also Eq.~\ref{nhcl}):

\beq
\mathcal{N}_{T,th}(\theta) = I_{th}(\Sigma_c,R_c) r_o \frac{1 - e^{-\epsilon\,tan\theta}}{sin\theta} \; , 
\label{}
\eeq

where

\beq
I_{th}(\Sigma_c,R_c)=k\pi\,\,R_{c,*}^{\alpha+3}\Sigma_{c,*}^{\beta+1}\Gamma(\alpha+3)\Gamma(\beta+1, x_{th}) \; .
\label{isrth}
\eeq

Here $x_{th} = \Sigma_{th}/\Sigma_{c,*}$ and $\Gamma(\xi, x_{th})=\int_{x_{th}}^{\infty}x^{\xi-1}e^{-x}dx$ is the incomplete Gamma function. 
The normalization $k$ of Eq.~\ref{isrth} can be expressed as a function of the total gas mass $M_{gas}$ and galaxy scale $r_o$ 
by considering Eqs.~\ref{rhoo} and \ref{totmgas}, and one gets:

\beq
\mathcal{N}_{T,th}(\theta) \sim 1.4 \frac{\Sigma_{gas}}{\Sigma_*} \frac{\mathcal{G}(\beta,x_{th})}{\beta} \frac{1 - e^{-\epsilon\,tan\theta}}{\epsilon\,sin\theta} \; , 
\label{nhclgasdist}
\eeq

where $\mathcal{G}(\beta,x_{th}) = \Gamma(\beta+1,x_{th})/\Gamma(\beta+1)$ decreases from 1 to 0 as $x_{th}$ increases from 0 to $+\infty$. Eq.~\ref{nhclgasdist} is the analog of Eq.~\ref{nhclgas}, which was derived for
a single cloud type. $\mathcal{N}_{T,th}(\theta)$ can be then estimated by specifying the total gas surface density and ``aspect ratio'' of the galaxy $\Sigma_{gas}$
 and $\epsilon$, respectively, and assuming that the surface densities of individual molecular clouds are distributed as a Schechter function of slope $\beta$ and characteristic density $\Sigma_{c,*}$. 

Hence, for a population of identical galaxies, the fraction of ISM-obscured nuclei with \nh~$>N_{\rm H,th}$ can be obtained as (see Eq.~\ref{fobsc}):

\beq
f_{obsc}(>N_{\rm H,th})=1-\int_{0}^{\pi / 2} e^{-\mathcal{N}_{T,th}(\theta)} \sin \theta d \theta \; .
\label{fnhth}
\eeq  

In Eq.~\ref{fnhth} one counts as obscured AGN with \nh~$>N_{\rm H,th}$ only those systems where individual clouds with  \nh~$>N_{\rm H,th}$ fall along the line of sight. Of course, more than one cloud with \nh~$<N_{\rm H,th}$ can lie along the line of sight and bring the total column density above the chosen threshold. Thus, the fractions of ISM-obscured AGN estimated from Eq.~\ref{fnhth} should be considered as lower limits. We nonetheless note that for a population of hig-$z$ disk galaxies observed at an average orientation angle of $\theta=57.3$ (see Sect.~\ref{elmet}), the average number of clouds along the line of sight is $\sim 1$ (see e.g. Fig.~\ref{nttheta}). We therefore consider Eq.~\ref{fnhth} as a good proxy for the ``true'' fraction of ISM-obscured AGN with \nh~$>N_{\rm H,th}$.

 \section{Results}\label{results}

 \subsection{Evolution of the ISM column density}\label{nhismz}

\begin{figure*}[t]
\begin{center}
\includegraphics[angle=0, width=11.0cm]{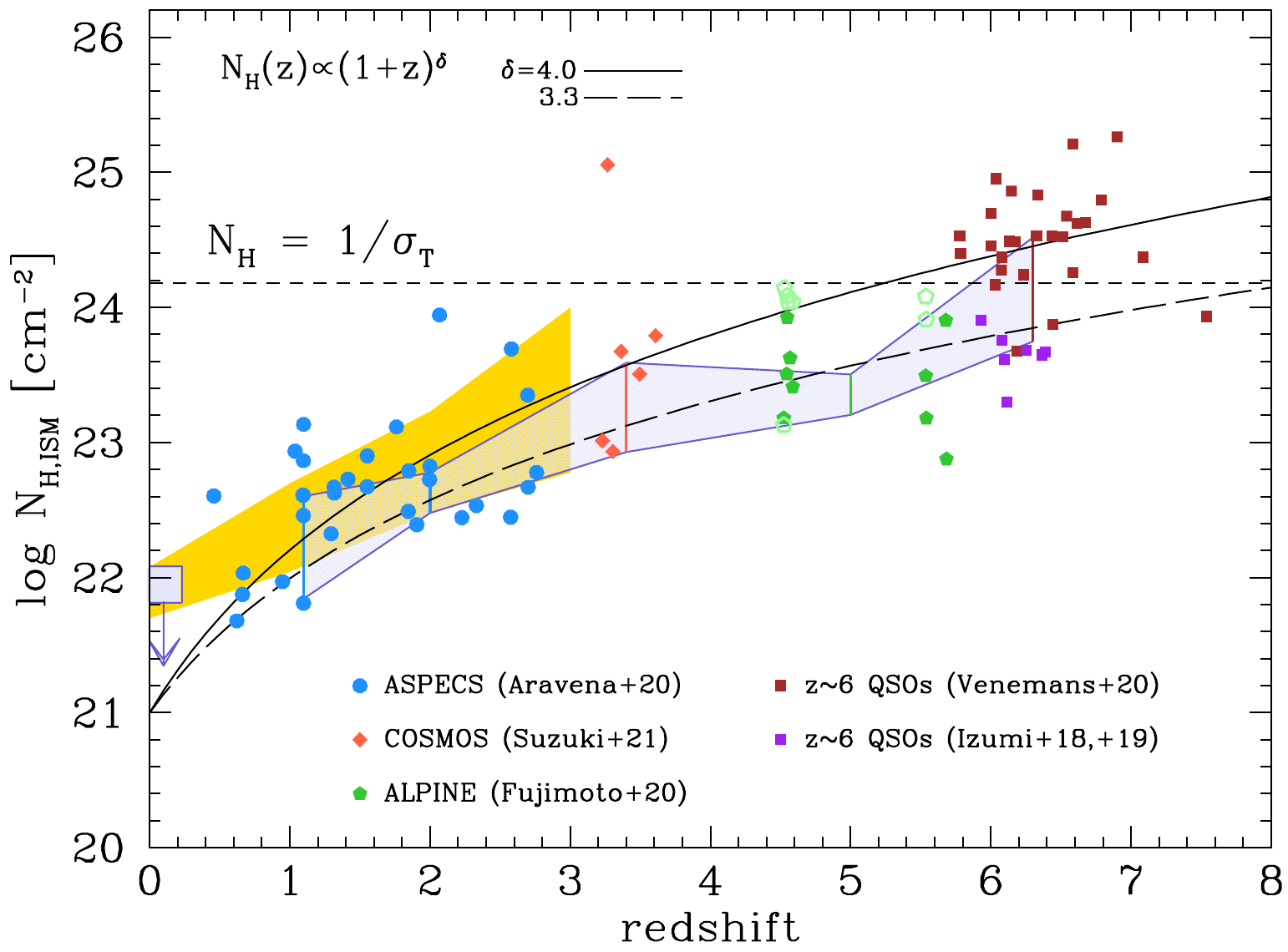}
\caption{ISM column density vs redshift for different samples of $M_*>10^{10}\msun$ galaxies as labeled. For ASPECS and COSMOS $z\sim3.3$ galaxies, ISM masses were derived
from dust masses, and ISM sizes from UV/optical stellar sizes. For ALPINE galaxies and for $z\sim6$ QSO hosts both ISM masses and sizes were derived from \cii\ data. For comparison, we also show for the ALPINE sample the results obtained using ISM masses derived from dust mass and ISM sizes derived from UV/optical stellar sizes (green open pentagons; see text for details). The upper bound of the light blue shaded area shows the median \nh\ values obtained for the different samples (ASPECS was divided in two redshift bins). The lower bound shows the median \nh\ values after including a conservative correction for incompleteness and observational biases (see Sect.~\ref{biases}). The \nh~$\propto(1+z)^4$ solid curve is obtained by combining empirical trends for the gas mass and galaxy size and is normalized to \nh$=10^{21}$\cm\ at $z=0$, which is consistent with the angle-averaged column density of the Milky Way \citep{willingale13} and with the upper limit to the median column density of local massive galaxies as derived from the EDGE-CALIFA survey (\citealt{bolatto17}; big square at $z=0.1$). The long-dashed curve running through the light-blue shaded area corresponds to $(1+z)^{3.3}$. The gold shaded area is from \citet{buchner17} as derived from the Illustris-1 simulation: lower and upper bounds are for galaxies with $M_*>10^{10}\msun$ and $M_*>10^{11}\msun$, respectively. The horizontal dashed line shows the column density corresponding to Compton-thick absorption.
} 
\label{nhz}
\end{center}
\end{figure*}

In Fig.~\ref{nhz} we show the distribution of the ISM column density as derived in Section~\ref{elmet} for the galaxy samples at different redshifts presented in Section~\ref{samples}. For ASPECS and COSMOS $z\sim 3.3$ galaxies,
we used the dust mass to estimate the total ISM mass and the UV/optical stellar size as a proxy for its size. For ALPINE and for the hosts of $z\sim6$ QSOs, we used \cii\ data for both gas mass and size estimates. 
For comparison, for the 7 continuum-detected sources in ALPINE with a reliable HST size estimate (see Sect.~\ref{sect:galsize}) we also computed ISM column densities using the dust mass and the UV/optical stellar size as proxies for the total ISM mass and size, respectively. On average, as shown in Fig.~\ref{nhz}, the column densities measured with this method are higher than those derived from \cii\ data (the median value is higher by $\sim 0.5$ dex). However, as discussed in Sect.~\ref{biases}, when accounting for selection biases, the  ranges allowed for the column densities derived with the two methods fully overlap.

A steady increase of the ISM column density with redshfit is observed, with typical values going from $\sim10^{22}$\cm\ at $z<1$ to $\sim10^{24}$\cm\ at $z>6$. That is, the ISM in early, massive galaxies can reach values as high as Compton-thick and hence provide significant obscuration to any nuclear emission. Interestingly, at $z\sim 6$, the ISM in the hosts of luminous QSOs \citep{venemans20} is denser than what is measured in the hosts of less luminous QSOs \citep{izumi18,izumi19}. Clearly, all known QSOs at $z\sim 6$ are UV-bright and, by selection, are seen along lines of sights free from such heavy ISM columns.  

By normalizing the ISM column density at $z=0$ to the average value observed in the Milky Way, \nh=$10^{21}$\cm\ \citep{willingale13}, the increase of the average ISM column density with redshift can be parameterized as \nh$(z)$/\nh$(0)\sim(1+z)^4$ (see black solid curve in Fig.~\ref{nhz}). The adopted value for the $z=0$ normalization is based on a single object and is then somewhat arbitrary. We nonetheless verified that this is consistent with what is measured in local samples of massive galaxies. To this end, we considered the 67 galaxies with $M_*>10^{10}\msun$ in the EDGE-CALIFA survey \citep{bolatto17}. Based on CO(1-0) observations with the CARMA interferometer \citep{bock06}, this survey measured the content and extent of the molecular gas reservoirs in a sample of nearby, $z<0.03$ galaxies that were detected at 22 $\mu$m by the Wide-field Infrared Survey Explorer (WISE; \citealt{wright10}). We used those measurements to derive a median surface density of $\Sigma_{gas}=4.3\times10^{21}$\cm, which translates (Eq.~\ref{nhave}) into a typical column density towards galaxy nuclei of \nh=$9.0\times10^{21}$\cm. Requiring a mid-IR detection with WISE pre-selects gas-rich galaxies. We then consider this column density estimate as an upper limit to the true, typical ISM obscuration in local massive galaxies,  and we plot it as a downward arrow in Fig.~\ref{nhz}.

Similarly, the overall ISM density vs redshift trend shown in Fig.~\ref{nhz} relies on samples of galaxies that are relatively bright in the FIR/mm domain, and one may then ask whether its validity is limited to those galaxy populations that are richer in gas at any redshift. We will show in the next Section that the ISM column density steeply increases with redshift even when accounting for selection biases, generalizing our results to the entire population of massive galaxies.

\begin{figure*}[h!]
\begin{center}
\includegraphics[angle=0, width=11.0cm]{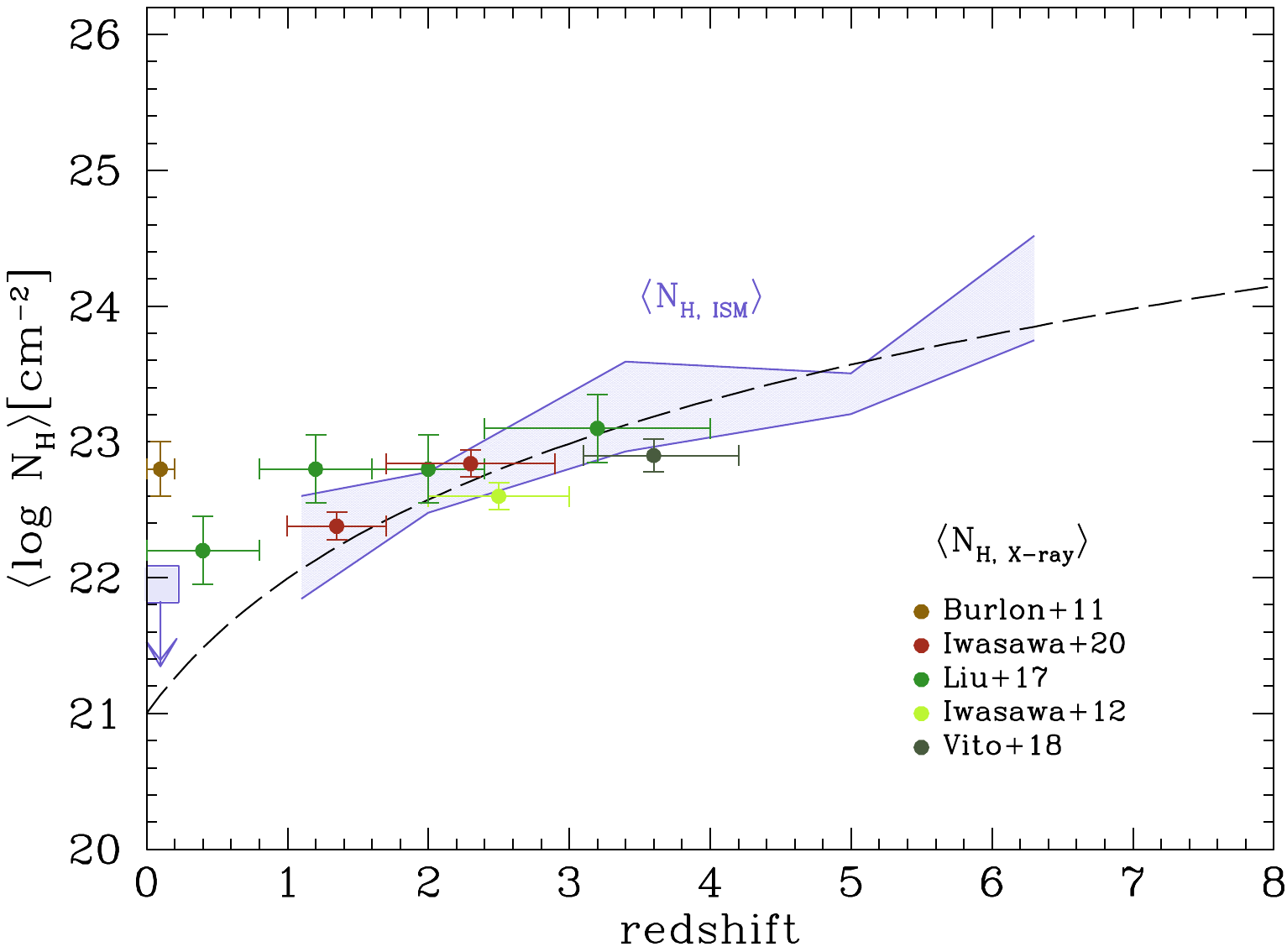}
\caption{Comparison between the evolution of the median column density of the ISM (light blue shaded area and black dashed curve -- same meaning as in Fig.~\ref{nhz}) and of the median absorption column density measured in different AGN samples through X-ray spectroscopy (see text for details). At $z\gtrsim1.5-2$ the ISM column density becomes comparable to that measured in the X-rays.} 
\label{nhzx}
\end{center}
\vspace{-0.2cm}
\end{figure*} 

 \subsection{Median \nhism\ and correction of selection biases}\label{biases}
 
By definition, any object with mm flux density below the ALMA detection limit is not included in the previous analysis. This can both bias high the overall $N_{\rm H,ISM}$ measurements and distort its redshift evolution. Broadly speaking, as the fraction of massive, gas poor galaxies increases with decreasing redshift \citep{calura08,fontana09}, one may expect that the strongest bias is at low-$z$. The strength of this bias is however obviously also related to the sensitivity of the considered ALMA observations. 

ASPECS is currently the deepest blind survey at 1.2~mm and is able to detect gas masses down to approximately $2.5\times 10^9\msun$ at any redshift above $z\sim0.5$. We then checked how many massive galaxies with $M_*>10^{10}\msun$ in the part of the HUDF covered by ASPECS ($\sim  2.1 \times 2.1$ arcmin$^2$) went undetected by ALMA. We considered the catalog of stellar masses computed by \citet{santini15} in the whole GOODS-S area, which also includes the HUDF and probes stellar masses down to $M_*\sim3\times 10^{7}\msun$ up to $z\sim3$.
\citet{santini15} compared several SED-based stellar mass estimates for GOODS-S galaxies as determined by different groups using the same photometry and redshift catalogs but different fitting codes, assumptions, priors, and parameter grids. In this work, we used the estimates flagged as ``13a$_{\tau}$'' in Santini et al. (2015; see their Table~1), which were derived assuming a Chabrier IMF and an exponentially decreasing star formation history, as they provide the best agreement with those published by \citet{aravena20} (we derived a null offset and $\sim0.25$ dex $rms$ for the mass ratio distribution of common objects). 
We matched the \citet{santini15} catalog with that of \citet{vanderwel12} and found 23 galaxies in the ASPECS survey area with reliable structural parameters ($q=0$ in \citealt{vanderwel12}), redshift in the range $z=0.5-3$ and stellar mass $M_*>10^{10}\msun$ (i.e. using the same cuts applied to the \citealt{aravena20} sample). Their size distribution is similar to that of ASPECS galaxies. We then used the size measured by \citet{vanderwel12} and assumed a gas mass of $2.5\times 10^9\msun$ to derive upper limits to their ISM column densities. Many of these limits fall in the cloud occupied by ASPECS galaxies
in Fig.~\ref{nhz}. However, in order to have a conservative estimate of the lowest possible median column density of the entire population of massive galaxies at $z=0.5-3$ (i.e. both detected and undetected by ALMA), we considered all of these 23 objects to lie below the lower envelope of the \nh\ vs $z$ distribution of ASPECS galaxies. By dividing the sample in two redshift bins, we found that, when ALMA undetected galaxies are included, 
the median $N_{\rm H,ISM}$ value decreases from $4\times 10^{22}$\cm\ to $7\times10^{21}$\cm\ at $z=1$, and from $6\times10^{22}$\cm\ to $3\times10^{22}$\cm\ at $z=2$ (see Fig.~\ref{nhz}). The quoted \nh\ ranges can therefore be considered as bracketing the true median values in both redshift bins. 

The six COSMOS $z\sim 3.3$ galaxies detected at 1.2~mm are drawn from the sample of \citet{suzuki21}, who observed with ALMA twelve star forming galaxies (11 with $M_*>10^{10}\msun$) with reliable gas phase metallicity measurements based on near-IR spectroscopy. Incidentally, by means of the dust-to-gas mass ratio $\delta_{GDR}$ vs gas-phase metallicity relation derived by \citet{magdis12}, \citet{suzuki21} estimated an average $\delta_{GDR}=210$ for their sample, in excellent agreement with what is assumed throughout this work ($\delta_{GDR}=200$). The median column density of the six ALMA-detected galaxies is log\nhism~$\sim23.6$. Assuming that the five ALMA-undetected massive galaxies in \citet{suzuki21} have gas masses and column densities smaller than ALMA-detected galaxies, the median ISM column density of the whole sample in \citet{suzuki21} would then coincide with the lowest column density measured, log\nhism~$\sim22.9$ (see Fig.~\ref{nhz}). All star forming galaxies in \citet{suzuki21} lie on the
main sequence. Since quiescent, gas-poor systems represent only a few percent of the whole population of massive galaxies at $z>3$ \citep{girelli19}, we consider the median \nhism\  estimated above as representative of the whole population.
  
Out of the 27 galaxies with $M_*>10^{10}\msun$ in the ALPINE sample of \citet{fujimoto20}, 21 were detected in \cii. The median column density of the nine galaxies with a reliable \cii-based size estimate is log\nhism~=23.5 (see Fig~\ref{nhz}, we plot it at $z=5$). We verified that the 12 galaxies with no good \cii-size estimate have a \nhism\ distribution similar to that shown in Fig.~\ref{nhz}, and we have assumed that the six \cii-undetected galaxies all have \nhism\ below the lowest measured value for the detected galaxies. This brings the corrected median log\nhism\ value of the whole sample to 23.2. Similarly to the COSMOS $z\sim 3.3$ sample, ALPINE galaxies lie on the main sequence \citep{fujimoto20}, and therefore
we consider the corrected median \nhism\ as representative of the whole population of massive galaxies at $z\sim 5$.   
We repeated the same exercise on the ALPINE sample by considering the ISM column densities derived from dust mass
and UV/optical sizes. The median column density of the 7 continuum detected galaxies with reliable HST size is log\nhism$\sim$24. When considering the whole sample of 13 dust-detected galaxies without requirements on the size reliability, we found that the lowest column density of the sample is about log\nhism$=$22.8. By conservatively assuming that the 14 continuum undetected galaxies all have column densities below this value, we estimate that a plausible range for the
median dust-based \nhism\ is 22.8-24, which is larger than the range derived from \cii\ data and fully encompasses it (see also Fig.~\ref{nhz}.) 

For the sample of $z\sim 6$ QSO hosts, we note that any correction to account for ALMA undetected objects does not affect significantly the mean \nhism\ derived from
ALMA detections only. As a matter of fact, all of the seven low-luminosity QSOs observed by \citet{izumi18,izumi19} are detected in \cii, and six of them also in the continuum.
The sample of high-luminosity QSOs in \citet{venemans20} instead includes only objects that were previously known to be FIR-bright. We evaluate the non-detection rate of high-luminosity QSOs hosts based on the sample discussed in \citet{venemans18} and \citet{decarli18}, which largely overlap with the \citet{venemans20} sample and include targets selected by UV luminosity only. The non-detection rate in that sample is less than $15\%$ for both \cii\ and continuum observations. As the ALMA observations in \citet{venemans20} are on average more sensitive than in \citet{venemans18}, this non-detection rate can be considered as an upper limit. Nonetheless, we checked that assuming a 15\% non-detection rate for luminous high-$z$ QSOs, would change the median \nhism\ estimates, either based on \cii\ or continuum emission, by less than 10\% and therefore we neglected this correction.
To derive a representative median \nhism\ for the entire population of $z\sim 6$ QSO hosts, one has instead to consider that the number of ALMA observations
of low-luminosity and high-luminosity QSOs do not reflect the actual abundance of the two populations. Bright QSOs were indeed discovered earlier and are easier targets, and then dominate the statistics of currently observed objects. As shown in Fig.~\ref{cfr_z6qso}, luminous QSOs are also the ones with larger gas masses and column densities, and therefore their median \nhism\ cannot be considered as representative of the whole population. To correct for this, we considered the UV-rest luminosity function (LF) of $z\sim6$ QSOs published by \citet{matsuoka19}, which combines the results from SHELLQs with those of shallower, wider-area optical surveys. By integrating the QSO LF in the range $M_{UV}=[-22,-25]$, i.e. the magnitude boundaries of the \citet{izumi18,izumi19} samples, and then at magnitudes $M_{UV}<-25$, as typical of the \citet{venemans20} sample, we estimate that low-luminosity QSOs are about 9 times more abundant than high-luminosity QSOs. We then smoothed the observed \nhism\ distributions of high- and low-luminosity QSOs with a boxcar of 1 dex in log\nhism, and resampled each distribution 1000 and 9000 times, respectively. The corrected median value of the final distribution is log\nhism=23.7, which we consider as representative of the observed population of $z>6$ QSOs, i.e. of $z\sim6$ massive galaxies. When compared with the allowed ranges for the median column density of massive galaxies at lower redshifts (see Fig.~\ref{nhz}), this calls for an increase of the median ISM column density with redshift as $(1+z)^\delta$, where $\delta$ is in the range 2.3-3.6. When the median column density of local galaxies is fixed to \nh=$10^{21}$\cm, as observed in the Milky Way \citep{willingale13}, the cosmic evolution of the corrected median ISM column density is best described by $\delta=3.3$ (see Fig.~\ref{nhz}). 

\subsection{Comparison with the median AGN X-ray column density}\label{sect:nhzx}

In Fig.~\ref{nhzx} we compare the evolution of the median ISM column density derived in the previous Section, both as measured and corrected for selection biases, with the median
column densities derived from X-ray observations of large AGN samples. To minimize the biases against heavily obscured AGN we considered either results from the deep \chandra\ and XMM-Newton observations of the CDFS \citep{iwasawa12, liu17, vito18, iwasawa20}, or, for the local Universe, results from AGN samples selected at energies above $\sim$10 keV with Swift/BAT \citep{burlon11} or NuSTAR \citep{zappacosta18}. 

As shown in Fig.~\ref{nhzx}, the median column densities derived for the X-ray nuclear obscuration are of the same order of the ISM column densities, especially 
at $z\gtrsim 1.5-2$. This indicates that the ISM may in fact produce a substantial fraction of the nuclear obscuration measured around distant SMBHs. Instead, in the local Universe, and likely up to $z\sim 1.5-2$, the median X-ray derived column densities appear in excess of what can be produced by the ISM, suggesting that most of the nuclear obscuration is 
produced on small scales by the AGN torus. 

We note that several caveats apply to the above comparison: first, the considered AGN samples generally do not contain any cut in host stellar mass, whereas our ISM density computations are valid for systems with $M_*>10^{10}\msun$. Most AGN in the considered X-ray samples, however, are likely hosted by massive galaxies. We checked that this is in fact the case e.g. for the BAT and for the \citet{vito18} samples, and that the average column densities obtained cutting at $M_*>10^{10}\msun$ are in good agreement with those of the total samples. Second, detecting the most heavily obscured, Compton-thick AGN is hard even for the deepest X-ray surveys \citep{lambrides20}. As a result, completeness corrections are difficult, and the median column density estimates may have been underestimated. Third, the fraction of obscured AGN is known to decrease with
X-ray luminosity \citep{gch07,h08} , which again may bias low the derived median \nh\ estimates, especially towards high redshifts where low-luminosity systems cannot be detected effectively. We defer to the Discussion a more detailed comparison between the X-ray and ISM obscuration. Here it is just worth remarking that, at high-$z$, the ISM column densities are of the same order of what is measured via X-ray absorption.

\subsection{Redshift evolution of the ISM-obscured AGN fraction}\label{sect:fobsc}

The formalism developed in Sections \ref{sub:clump},\ref{sub:clump2},\ref{sub:clump3}, can be used to follow the redshift evolution of the average number of ISM clouds towards galaxy nuclei and, in turn, of the ISM-obscured AGN fraction. Since the surface density of individual molecular clouds has been found to significantly increase with redshift for a given cloud size \citep{mirka19}, we allow the characteristic surface density $\Sigma_{c,*}$ of molecular clouds to increase with redshift as $\Sigma_{c,*}(z)=\Sigma_{c,*}(0)(1+z)^\gamma$, exploring the range $\gamma=2-3$ to account for the large uncertainties associated with the above trend. We instead assume for simplicity that the distribution of the cloud radii does not vary with redshift, as observations of molecular clouds in distant galaxies are unavoidably biased against small systems and hence cannot probe the entire size distribution. As an example, the ALMA observations of the ``Snake'' lensed galaxy at $z=1$ in \citet{mirka19} could only resolve clouds with $R_c>30$pc.

The redshift evolution of the ISM-obscured AGN fraction above a given \nhism\ threshold is derived using  Eqs.~\ref{nhclgasdist} and \ref{fnhth}. The redshift dependence is encapsulated in the function 
$(\Sigma_{gas}/\Sigma_{c,*})\times \mathcal{G}(\beta,x_{th}) \propto(1+z)^{\delta-\gamma}\mathcal{G}(\beta,x_{th}(z,\gamma))$. Once the characteristic gas surface density and
cloud surface density distribution of galaxies at z=0 are specified (that is, $\Sigma_{gas}(0)$, $\Sigma_{c,*}(0)$, and $\beta$), and assuming that the ratio, $ \epsilon$, between the number of galaxy clouds
seen face-on vs edge-on is the same at all redshifts, the cosmic evolution of the ISM-obscured AGN fraction is determined by how fast the
galaxy total gas surface density increases with respect to  the characteristic cloud surface density $(1+z)^{\delta-\gamma}$ modulated by $\mathcal{G}(\beta,x_{th}(z,\gamma))$. Now $\mathcal{G}$ depends on $z$
and $\gamma$ through the lower integration limit $x_{th} = \Sigma_{th}/(\Sigma_{c,*}(1+z)^\gamma)$: for a fixed $N_H$ ($\Sigma$) threshold, it increases with increasing redshift at a rate that grows with increasing $\gamma$. 

\begin{figure}[t]
\includegraphics[angle=0, width=9cm]{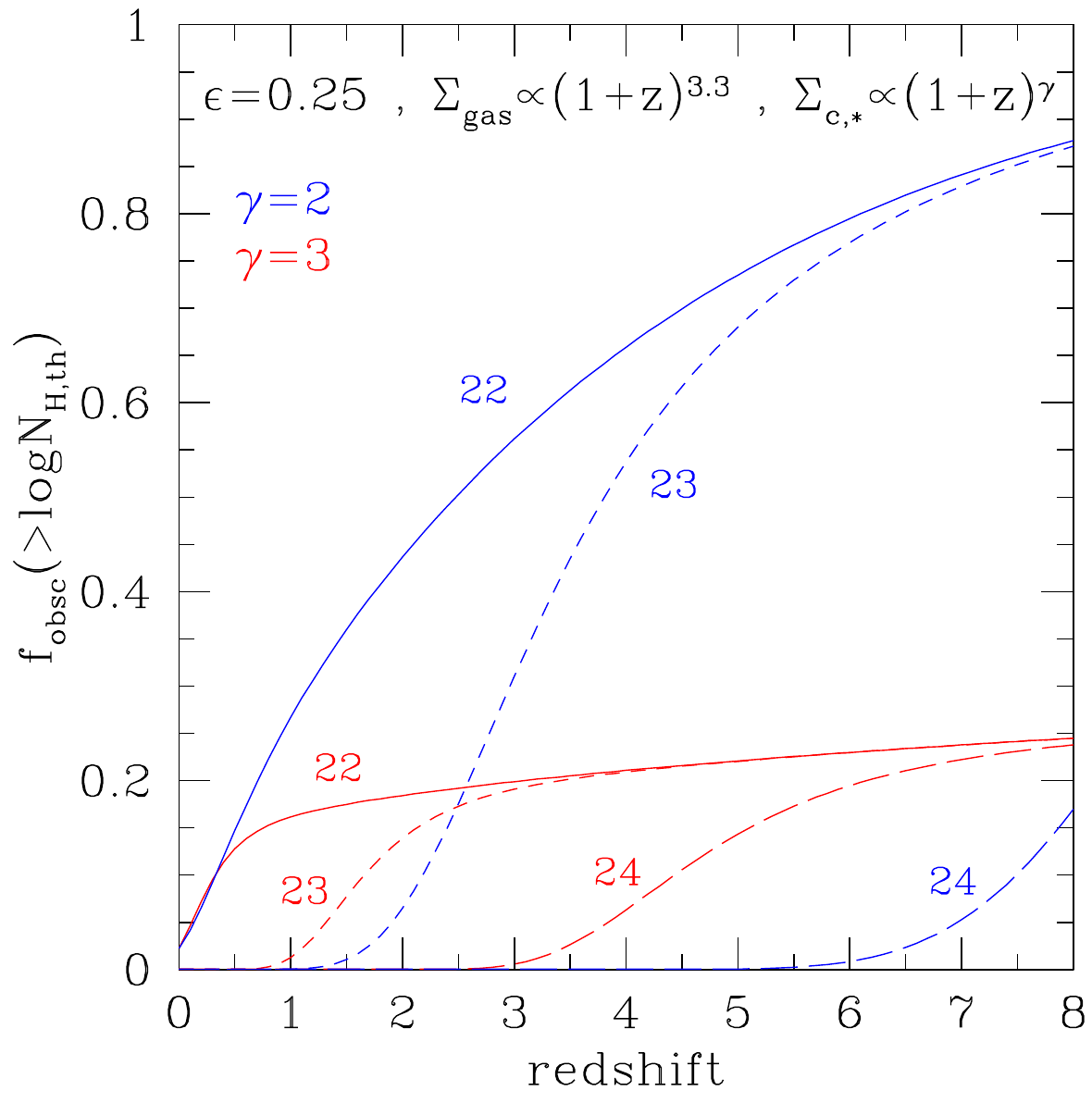}
\caption{Expected redshift evolution of the ISM-obscured AGN fraction. Blue curves consider a slower ($\gamma=2$) increase with redshift of the characteristic cloud density $\Sigma_{c,*}$ . Red curves assume a faster increase ($\gamma=3$). Labels refer to different thresholds in log column density (see text for details).} 
\label{fobscfig}
\end{figure}

We show in Fig.~\ref{fobscfig} the cosmic evolution of the fraction of ISM-obscured AGN  for $\gamma$=2 and 3, and for three different column density thresholds, $N_{th}=10^{22}$, $10^{23}$, and $10^{24}$\cm. Based on the evolution of the ISM column density (Section~\ref{biases}), which is linked to the total gas surface density through Eq.~\ref{nhave}, we assume $\delta=3.3$ to describe the increase in the total gas surface density with redshift.
We also fixed the total gas surface density of local galaxies to $\Sigma_{obs}(0)\sim 5\times10^{20}$\cm, corresponding to an average column density
towards the nucleus of \nh$\sim 10^{21}$\cm, and the shape parameters of the cloud density distribution to what is observed for the inner Milky Way clouds ($\Sigma_{c,*}(0)=25$\msunpc, $\beta=1.8$).
We also fixed $\epsilon=0.25$, consistently with the adopted value of $h/r_e=0.15$ assumed in Section ~\ref{elmet}.

As shown in Fig.~\ref{fobscfig}, given the global increase of the ISM column density with redshift, the cosmic evolution of the ISM-obscured AGN fraction heavily depends on the evolution of the characteristic cloud density $\Sigma_{c,*}$: on the one hand, if $\Sigma_{c,*}$ rapidly grows with redshift ($\gamma=3$), the fraction of ISM-obscured AGN evolves weakly, as only a few, dense clouds are needed to match the total column density increase with redshift. On the other hand, if $\Sigma_{c,*}$ grows slowly ($\gamma=2$), more clouds are needed to produce the same column, and hence there are more chances to see the nucleus through a cloud. In both cases, at $z=0$, the fraction of ISM-obscured AGN  is expected to be below 
5\% for a column density threshold of $N_{H,th}=10^{22}$\cm, and to be negligible for $N_{H,th}=10^{23}$\cm (see Fig.~\ref{fobscfig}). At $z=3.5$, instead, $f_{obsc}(>10^{23})$ grows from 0.21 to 0.43 when moving from $\gamma=3$ to $\gamma=2$. As expected, the fraction of heavily obscured nuclei $f_{obsc}(>10^{23})$ is always smaller than what is obtained by relaxing the absorption threshold $f_{obsc}(>10^{22})$, but the two numbers progressively get closer towards higher redshifts as the fraction of low-column clouds in galaxies gets smaller. Remarkably, in the $\gamma=2$ scenario, more than 80\% of all early SMBHs at $z>6$ may be obscured by the ISM in their hosts. At even higher redshifts,  $z>8$, one may further expect to see Compton-thick absorption from the ISM in about 15-20\% of accreting SMBHs.
Clearly, the cosmic evolution of the surface density of individual molecular clouds is largely unknown. On the one hand, by comparing the typical densities of clouds in the ‘Snake’ galaxy at $z\sim 1$ \citep{mirka19} with that of Milky Way clouds one would get $\gamma>3$ (which may be however biased high as selection effects would favor the detection of bigger and denser clouds in the `Snake'). On the other hand, theory works on $z\sim6$ galaxies (e.g. \citealt{vallini18,vallini19}) often assume fiducial cloud sizes and masses corresponding to surface densities similar to what is measured in the Milky Way and other nearby star forming galaxies, meaning $\gamma=0$. We actually do not consider the $\gamma=2$ scenario as the more plausible {\it a priori}, but anticipate that it provides the best representation of the obscured AGN fractions measured at different cosmic times (see Sect.~\ref{cotot}).

\begin{figure}[t]
\includegraphics[angle=0, width=9cm]{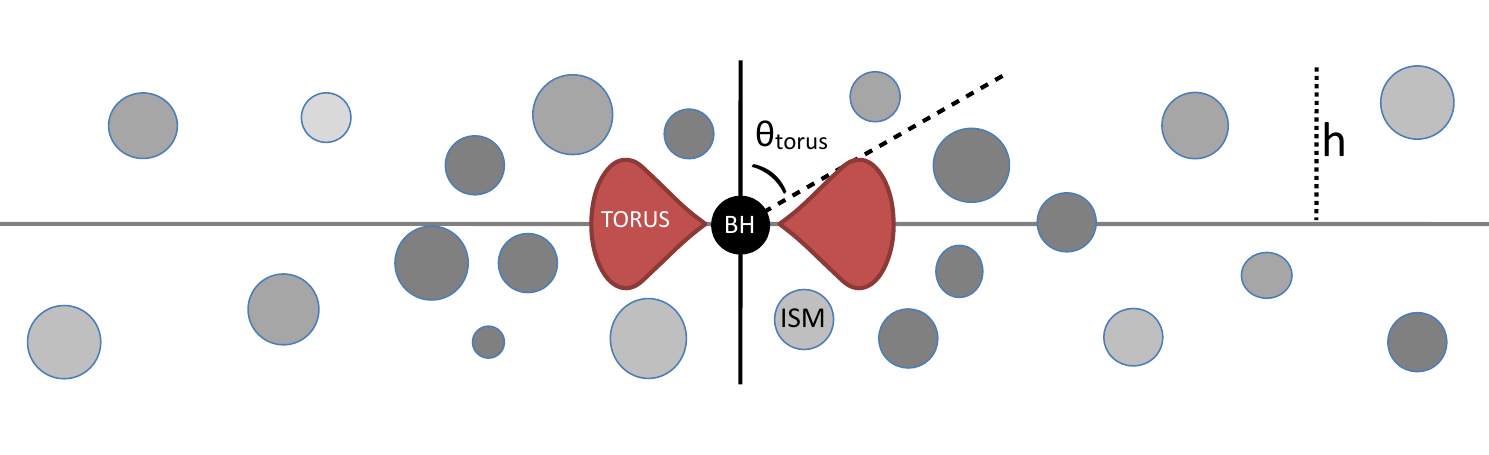}
\caption{Edge-on view of the considered geometry for the ISM clouds and the torus (not to scale).} 
\label{fig:scheme_clouds}
\end{figure}

\begin{figure*}
\includegraphics[angle=0, width=\textwidth]{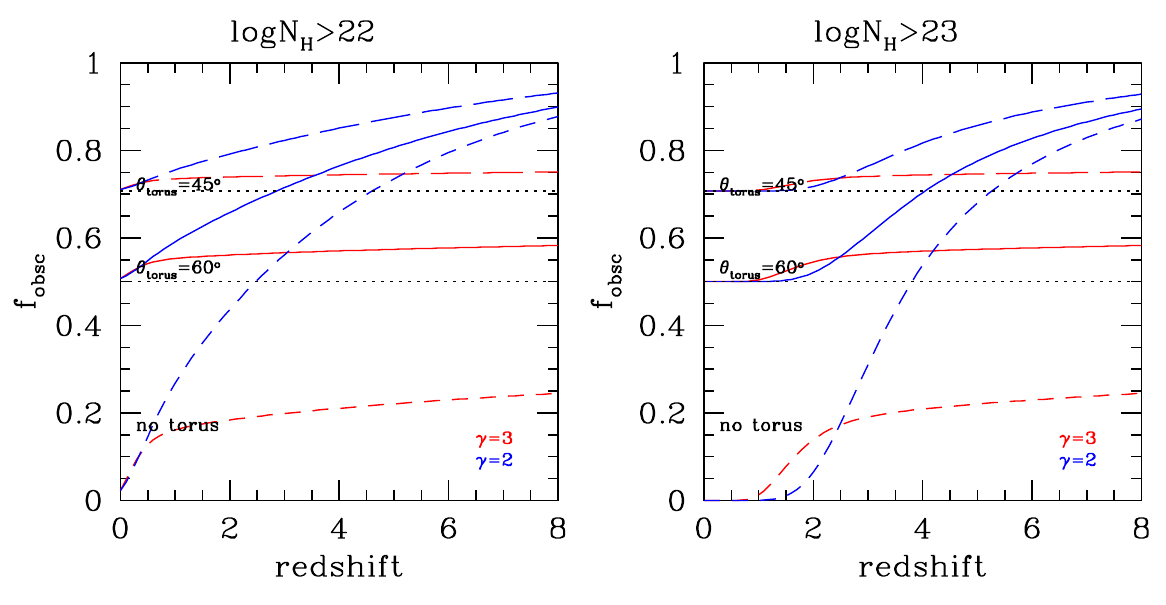}
\caption{Total fraction of obscured AGN with log$N_H>22$ ({\it left panel}) and log$N_H>23$ ({\it right panel}) vs redshift for different combinations of large-scale (ISM) and small-scale (torus) obscuration. All curves are for $\delta=3.3$ and $\epsilon=0.25$. The characteristic surface density $\Sigma_{c,*}$ of ISM clouds is assumed to evolve as $\propto(1+z)^\gamma$, with $\gamma=2$ (blue curves) or $\gamma=3$ (red curves). Short dashed lines refer to ISM-only obscuration (i.e. no torus considered, as in Fig.~\ref{fobscfig}). Long dashed and solid lines show the total obscured AGN fraction when adding a torus with half-opening angle $\theta_{torus}=45^o$ and $60^o$, respectively. The horizontal dotted lines show the obscured AGN fraction expected by torus-only obscuration.}
\label{fobsc2_curves}
\end{figure*} 

\begin{figure*}
\includegraphics[angle=0, width=\textwidth]{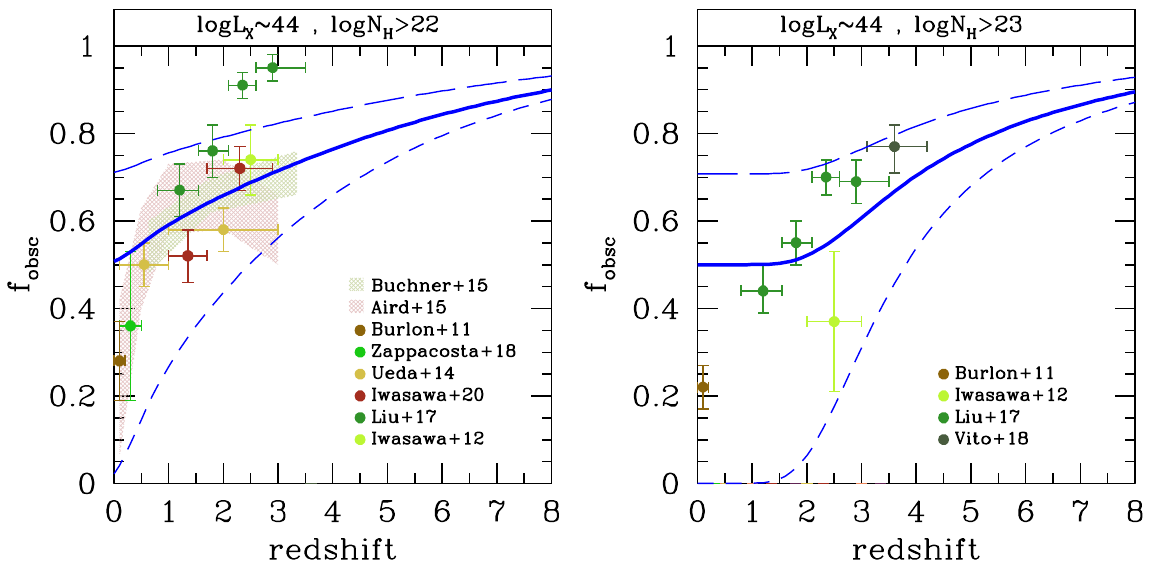}
\caption{Total fraction of obscured AGN with log$N_H>22$ ({\it left panel}) and log$N_H>23$ ({\it right panel}) vs redshift measured in different X-ray samples (see labels) and compared with model curves of large-scale (ISM) plus small-scale (torus) obscuration. Only samples with median log$L_X\approx44$ have been considered here (see text for details). Curves have the same meaning as in Fig.~\ref{fobsc2_curves}. The model with $\gamma=2$ and $\theta_{torus}=60^o$ (thick solid blue line) provides a good representation of the data.} 
\label{fobsc2_data}
\end{figure*} 

\begin{figure}
\includegraphics[angle=0, width=0.49\textwidth]{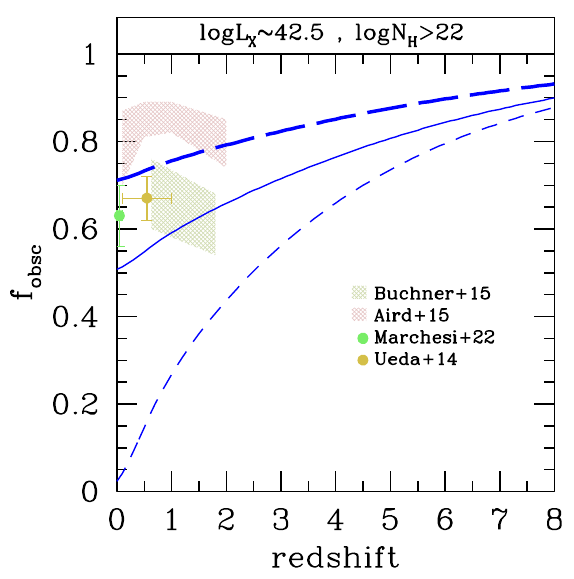}
\caption{Total fraction of obscured AGN with log$N_H>22$ vs redshift observed in different X-ray samples of low-luminosity AGN (log$L_X\approx42.5$; see text for details). Model curves are as in Fig.~\ref{fobsc2_data}. The model with $\gamma=2$ and $\theta_{torus}=45^o$ (thick dashed blue line) provides a reasonable representation of the data.} 
\label{fobsc2_lowlum}
\end{figure} 

\section{Discussion}\label{discuss}

\subsection{ISM column density: uncertainties and empirical trends}

In Section~\ref{biases}, we showed that the median column density of massive galaxies increases with redshift as \nh\ $\propto (1+z)^{\delta}$, where $\delta=3.3$. The exact value of the slope $\delta$ is admittedly uncertain, and an adequate representation of the median column density values observed at different redshifts would still be obtained for $\delta=2.3-3.6$. Nonetheless, the derived slope is similar to what one would derive from simple empirical trends reported in the literature about the cosmic evolution of the average gas mass and size of massive galaxies. The gas mass of galaxies is in fact known to rapidly increase with redshift as $(1+z)^a$, with $a\sim2$ \citep{scoville17,tacconi18,liu19}. For instance, for main sequence star forming galaxies, \citet{scoville17} found that the ISM content increases as a function of redshift as $M_{ISM}\propto(1+z)^{1.84\pm0.14}$, and is also a weak function of stellar mass ($\propto M_*^{0.30\pm0.04}$, see their Eq.6). We note that the \citet{scoville17} sample is essentially made by galaxies with stellar masses in the same range of the previously considered ASPECS sample, $M_*=1-30 \times 10^{10}\;M_{\odot}$ (see Sect.~\ref{aspecs}): in fact, less than 3\% of the $>700$ objects  in \citet{scoville17} have masses below $10^{10}\;M_{\odot}$ and 1\% have masses above $3\times10^{11}\;M_{\odot}$. 
The size of galaxies is known to decrease with redshift (e.g. \citealt{allen17}). This decrease is commonly parameterized as $r_e\propto(1+z)^{-b}$, where $r_e$ is the effective (or half-light) radius of the galaxy and $b\sim1.0$. The exact value of $b$ depends on the specifics of the galaxy population under consideration. For instance, \citet{shibuya16} and \citet{shibuya19} measured $b=1.20$ and 1.37 for LBGs and LAEs, respectively. For a mass-complete sample of star forming galaxies with $M_*>10^{10}\;M_{\odot}$ in the redshift range $z=1-7$ selected in the CANDELS fields, \citet{allen17} found $b$ in the range 0.9-1.0, depending on the adopted IMF and analysis methodology. They measured the effective radius by means of HST F160W photometry. The upper mass bound of the \citet{allen17} sample is again $M_*\sim3\times10^{11}\;M_{\odot}$. Assuming that gas and stars follow the same spatial distribution and then have equal effective radii, by combining the empirical relations found by \citet{scoville17} and \citet{allen17} one would trivially get that the average ISM volume and surface density increases with redshift as $(1+z)^5$ and $(1+z)^4$, respectively, for galaxies within the same stellar mass range considered in this work. The slope of the column density relation, $\delta=4$, derived from this simple exercise is then similar to what we derive in Sect.~\ref{biases} from our samples ($\delta=3.3$), and again points to a steep increase of the ISM density towards early epochs.

Concerns in the evolution rates derived for the ISM density may be related to how gas masses and sizes have been estimated. As discussed in Sect.~\ref{sect:galsize}, the effective radii of stellar emission as tracked by UV/optical rest-frame data are similar to those derived from CO emission, but a factor of 1.5-1.7 larger than those measured from dust continuum emission \citep{fujimoto17,lang19}. 
\citet{calistro18} suggest that this difference is due to temperature and optical depth gradients which cause a steeper drop in the FIR emission at large radii: stellar emission would then be a more reliable tracer of the spatial distribution of cold molecular gas, as also found by recent numerical simulations \citep{popping22}. This seems to be true at all redshifts, as the effective radius measured from ALMA continuum emission decreases with redshift with the same rate ($b\sim 1$) as measured from CANDELS (UV/optical-rest) photometry \citep{fujimoto17}.
In Sect.~\ref{sect:galsize} we used for ASPECS and COSMOS $z\sim 3.3$ galaxies the stellar sizes derived from HST photometry. For the $z>4$ samples, ALPINE and $z\sim 6$ QSOs,  
\cii-based size measurements are available and we resorted to those. For the ALPINE sample, the \cii-based sizes are $\sim2\times$ larger than the rest-frame UV-continuum size
measured with HST \citep{fujimoto20} We note that, as yet, no public information is available on the dust-continuum sizes for this sample. For the high-$z$ QSOs samples, \cii-based sizes are on average $\sim1.5\times$ larger than those estimated from the dust continuum (the host UV/optical stellar emission is as yet undetected in these systems). Hence, in general, we may conclude that: i) using dust-based measurements would likely produce a global increase of the ISM densities by a factor of $\sim2-4$ at all redshifts, making ISM-obscuration of AGN even stronger; ii) replacing the \cii\ sizes with the optical/UV-rest sizes (when available) would increase by a factor of $\sim 4$ the ISM densities of our highest redshifts samples, hence making the observed evolution even steeper.

\subsection{Adding the pc-scale torus}\label{addtor}

In Sect.~\ref{sect:fobsc} we discussed how the redshift evolution of the ISM-obscured AGN fraction depends on our cloud model assumptions, particularly on the evolution
of the cloud characteristic density $\Sigma_{c,*}$. Fig.~\ref{fobscfig} shows that, if the evolution in  $\Sigma_{c,*}$ is not too fast, the fraction of ISM-obscured AGN rapidly increases towards early cosmic epochs. The model predictions, however, cannot be directly compared with the observed fractions of X-ray obscured AGN, as these include any absorption along the line of sight, especially that produced by dense, dusty gas around the nucleus.This so-called ``torus'', is thought to be azimuthally distributed within a few to tens of parsecs from the SMBH~\citep[e.g.][]{combes19,garcia19}. Locally, the torus covering factor is used to explain the abundance of heavily obscured SMBHs with $N_H>10^{23}$\cm \citep{risaliti99,marchesi18}, as opposed to the local global ISM, which, at $z\sim0$, can only produce mild obscuration (\nhism$\lesssim 10^{22}$\cm ; see Sect.~\ref{nhismz} and Fig.~\ref{nhz}).

Recent progress has been made in both observations \citep{zhao20,torres21} and modeling \citep{balokovic18,buchner19} of the torus component. Accurate fits to the NuSTAR and XMM-Netwon spectra of local Seyfert 2 galaxies \citep{marchesi19,zhao20} revealed that the torus covering factor is about 0.6-0.7. Although the torus structure is likely complex and patchy, in the simple assumption of a medium uniformly distributed in a donut-like shape around the SMBH, a covering factor of 0.71 corresponds to a half-opening angle of $\theta_{torus}=45^o$. Moving from Seyfert to QSO luminosities, the torus covering factor is known to decrease, possibly because of the increased radiation pressure \citep{maiolino07,lusso13,ricci17}: at $L_{bol}\sim 10^{46-47}\lsun$, the torus covering factor is estimated to be 0.4-0.5, where 0.5 corresponds to a half-opening angle of 60$^o$. Here we shall now assume that the torus covering factor does not evolve with redshift. This is supported by the lack of evolution in the broad band SEDs of luminous quasars, which are remarkably self-similar up to the  highest redshifts, $z\sim6-7$ \citep{barnett15,nanni18}. In particular, no evolution has been observed in the typical ratio $\mathcal{R}$ between the QSO UV/optical and mid-IR luminosity \citep{leipski14,bianchini19}, which is considered as a proxy for the fraction of nuclear radiation reprocessed by hot dust ($T<1500$~K) within a few parsecs from the nucleus, i.e. for the torus covering factor. For example, \citet{bianchini19} divided a sample of $>250,000$ luminous QSOs at $1<z<5$ in three redshift bins populated by objects with similar luminosities\footnote{The average QSO luminosity varies by less than a factor of 2 from the lowest to the highest redshift bin} and found a remarkably constant average value $\mathcal{R}\sim1.1$ in the three bins. This suggests that the covering factor of the small-scale, circumnuclear medium is primarily regulated by QSO feedback across cosmic times. 

ALMA observations of nearby Seyfert galaxies at pc-scale resolution showed that the torus axis and the galaxy disk axis are often misaligned \citep{combes19}. Accounting for any torus-disk misalignment in our model (see below) would increase the total gas covering factor around the nucleus, and, in turn, the total obscured fraction of AGN at any redshift. For simplicity, we considered the torus and the galaxy disk as co-axial (see Fig.~\ref{fig:scheme_clouds}), keeping in mind that the derived obscured AGN fractions may represent lower limits. Again, for the sake of simplicity, we assumed that the torus produces Compton-thick absorption through all lines of sights crossing it. We considered two cases for its half-opening angle: $\theta_{torus}=45^o$ or 60$^o$.  In this simple configuration, the total fraction of obscured AGN can be simply obtained with Eq.~\ref{fnhth} by using $\theta_{torus}$ instead of 90$^o$ as the integral upper bound, as all lines of sights with $\theta_{torus}<\theta<90^o$ are now heavily obscured by the torus component.

\subsection{Cosmic evolution of the total nuclear obscuration}\label{cotot}

In Fig.~\ref{fobsc2_curves} we show the effects of adding the small-scale torus absorption to the large-scale ISM absorption for different assumptions on the evolution of the
characteristic cloud surface density. As in Fig.~\ref{fobscfig}, we fixed the evolution rate of the average ISM density to $\delta=3.3$ and the ratio between the average number of
ISM clouds crossed face-on vs edge-on to $\epsilon=0.25$. If the evolution in $\Sigma_{c,*}$ is fast, $\gamma=3$, the torus covering factor overwhelms that of ISM clouds. In this case, the obscured AGN fraction is expected to mildly increase up to $z\approx 1-2$ and then stay almost constant. For $\gamma=2$ instead, the rapid increase in the ISM covering factor with redshift is not washed out by the constant torus coverage, and an overall increase of the obscured AGN fraction towards early cosmic times is then expected. 
In Fig.~\ref{fobsc2_data} we compare the model curves for $\gamma=2$ with the obscured AGN fractions measured in different X-ray samples for two different $N_H$ thresholds ($10^{22}$ and $10^{23}$\cm). In addition to the X-ray samples already used in Section~\ref{sect:nhzx}, we also considered the results from XLF-based computations that try to correct for selection biases against heavily obscured sources \citep{ueda14,aird15,buchner15}. As the fraction of obscured AGN depends on X-ray luminosity \citep{gch07,h08} (see also the discussion above), we considered separately high-luminosity and low-luminosity AGN, using samples with intrinsic, median 2-10 keV luminosity log$L_X=44\pm0.3$ \ergs for the high-luminosity bin, or log$L_X<43$ \ergs for the low-luminosity bin.


The ISM model with $\gamma=2$, combined with a torus with $\theta_{torus}=60^o$ (the expected opening angle at log$L_X\sim 44$; see above), is able to explain quantitatively the increase with cosmic time in the fractions of luminous AGN with $N_H>10^{22}$\cm\ and $N_H>10^{23}$\cm\ (see solid lines in Fig.~\ref{fobsc2_data}). In Fig.~\ref{fobsc2_lowlum} we  show the same comparison for obscured AGN of lower luminosities  (log$L_X\sim42.5$~\ergs). Here we also considered a recent update on the fraction of obscured 
AGN measured locally that is based on the BAT 100-month sample (Marchesi et al. in prep). At these low luminosities, measurements of the obscured AGN fraction are clearly sparser and incomplete, especially towards high redshifts, but the same model of ISM plus torus absorption (this time using a torus opening angle $\theta_{torus}=45^o$, as expected for low-luminosity AGN) provides a good match to the data. 

Our results are subject to a number of uncertainties, both in the data and in the model, as discussed below. In Sect.~\ref{sect:nhzx}, we mentioned that no pre-selection in their host stellar mass was originally applied to the X-ray AGN samples considered here. Nonetheless, we expect that the obscured AGN fractions would not change significantly when cutting the sample at $M_*>10^{10}\msun$, as most AGN in those samples are likely hosted by massive galaxies. Also, because of the difficulties to detect Compton-thick AGN in sizable numbers, many studies just measure the fraction of obscured objects within the Compton-thin AGN population (log$N_H<24$). Assuming that the relative abundance of Compton-thick AGN is similar to what is assumed in the CXB synthesis model of \citet{gch07}, we estimate that a plausible correction for the missing Compton-thick AGN would increase the total fractions of obscured AGN by $\Delta f_{obsc}\sim 0.1$ on average. Finally, although the observed increase in the obscured AGN fraction may look somewhat steeper than the expectations (Fig.~\ref{fobsc2_data}), we note that the model curves are not meant to best-fit the data and may be adjusted by varying the input parameters. For instance, one could modify the face-on vs edge-on cloud number ratio $\epsilon$, or introduce different density gradients in the cloud distribution along the galaxy disk thickness, to cope with the variations in the galaxy morphology across cosmic time, as the fraction of mergers and irregular systems is known to increase with redshift \citep{mortlock13,whitney21}. In addition, one cannot entirely rule out a redshift evolution of the torus covering factor, albeit this might be at odds with the lack of evolution in the UV/optical to mid-IR luminosity ratio observed in luminous QSOs  at different redshifts \citep{bianchini19} and discussed in Sect.~\ref{addtor}. For example, the observed increase in the obscured AGN fractions shown in Fig.~\ref{fobsc2_data} could be also reproduced assuming that the fraction of ISM-obscured AGN does not evolve significantly (i.e. $\gamma=3$), but that the torus half-opening angle evolves from $\theta_{torus}=60^o$ at $z=0$ to $\theta_{torus}=45^o$ at $z\sim 4$ (see e.g. Fig.~\ref{fobsc2_curves}). Exploring the full model parameter space is, however, beyond the scope of this paper and is deferred to future work. With all this in mind, we believe that our combined, ISM plus torus model provides a good description of the observed trends: obscuration by the ISM alone would be too small to explain local data; torus obscuration only would not explain the observed increase towards early cosmic times. 

Interestingly, such a model predicts that the fraction of obscured AGN is about 80-85\% at $z=6$ and $>90$\% at $z>8$ ($\sim$50-55\% of the systems being Compton-thick). That is, the bulk of SMBHs in the first 0.5-1 Gyr of the Universe are expected to be hidden by dust and gas distributed across all galaxy scales, from pc to kpc. 
Currently, we do not have any known example of such obscured systems. The best candidates to date may hide among the population of $z\sim6$ LBGs and QSOs discovered by the Subaru HSC surveys. Among the published 93 SHELLQs QSOs, 18 objects in fact exhibit narrow Ly$\alpha$ with FWHM $<$ 500 km~s$^{-1}$. Since the Ly$\alpha$ luminosity exceeds $10^{43}$\ergs, as in AGN-dominated systems (\citealt{sobral18}), they are tentatively classified as QSOs. The origin of these narrow line (NL) objects is, however, not entirely clear, since star forming galaxies could produce strong Ly$\alpha$ and high-ionization condition reminiscent to AGN \citep{nakajima18a}. Deep X-ray observations will be a viable method to reveal whether obscured QSOs really hide among them.

\subsection{Comparison with other works}

\citet{buchner17} used X-ray spectra of long GRB afterglows to derive ISM column densities in a sample of galaxies in the redshift range 0.5-3 with $>$90\% redshift completeness and selected independently of their stellar mass (SHOALS sample; \citealt{perley16}). They found a positive correlation between ISM column density
and galaxy stellar mass of the form \nhism\ $\propto M_*^{1/3}$. The median column density of their high-mass sample with $M_*>10^{10}\msun$, is log\nhism$=22.13\pm0.10$. This value was derived using GRB at random locations along galaxy disks, and would increase to log\nhism$\sim22.4$ (Eq.~\ref{nhave}) when seen from galaxy nuclei. This is already
in agreement within a factor of $\sim 2$ with the median ISM column density towards the nuclei of ASPECS galaxies, cut at the same mass threshold, and with a similar redshift range
(log\nhism$=22.70\pm0.10$), but the agreement may further improve by considering that: i) as reported in \citet{perley16}, the GRB rate per unit star formation in the SHOALS sample is constant in galaxies with gas-phase metallicity below the solar value but heavily suppressed in more metal-rich environments. Therefore, the most massive and gas-rich galaxies may be under-represented in GRB-selected samples; ii) as discussed in Sect.~\ref{biases}, the ASPECS sample only includes gas-rich galaxies detected by ALMA with $M_{gas}\gtrsim 3\times 10^9\msun$ and, when accounting for ALMA-undetected galaxies, the 'true' average column density of massive galaxies at $z=0.5-3$ would likely decrease by a factor of $\approx 2$. In summary, our ISM column density estimates are in good agreement with those of \citet{buchner17} for galaxies with similar stellar masses and redshifts. However, as opposed to \citet{buchner17}, we found evidence that \nhism\ evolves with redshift even within the ASPECS sample, with typical values increasing by more than one dex from $z=0.5$ to $z=3$ (see e.g. Fig.~\ref{nhz}). 

Recently, \citet{masoura21} analyzed the host galaxy properties of $>$300 X-ray selected AGN in the XMM-XXL survey \citep{pierre16}, without finding any strong correlation between 
AGN type (obscured vs unobscured) and host physical properties. In particular, they found no correlation between the nuclear column density as estimated from X-ray colors (hardness ratios, HR) and star formation rate, and hence suggested that the obscuration is not related to the large scale star formation in the galaxy, i.e. to the host ISM. We note that our findings are not in contrast with those of \citet{masoura21}. In fact, in our model both obscured and unobscured AGN share the same ISM content, hence the same SFR, and the AGN type classification depends only on the chance that the line of sight to the nucleus is occulted by an ISM cloud. It is then the fraction of obscured AGN, rather than the AGN average column density, that is the quantity that we expect to correlate more with SFR in a given population of galaxies.

\subsection{Comparison with theory}


In Fig.~\ref{nhz} we compared the redshift evolution measured for the ISM density with the expectations extracted by \citet{buchner17} from the Illustris-1 numerical simulation \citep{vogelsberger14}, which reaches spatial resolutions down to $\sim50$~pc. By using a ray tracing technique, \citet{buchner17} derived the average
column density of ISM metals in simulated galaxies with different stellar masses and in different redshift bins. The metal column density was then converted to total ISM column density assuming solar abundances. Based on their Fig.~16, we inferred the evolution of the average ISM column density in the redshift range $z=0-3$ at two fixed stellar masses, $10^{10}$ and $10^{11}\msun$. Here we do not apply the factor of $\sim2$ correction to go from average to nuclear column density (as per Eq.~\ref{nhave}) since \citet{buchner17} perform their
ray-tracing analysis starting from the densest regions of simulated galaxies, which are presumably those at galaxy centers. The agreement between the measured and the expected evolution of the ISM density of massive galaxies is very good in the range $z\sim1-3$. At lower redshifts, the simulated trend is just consistent with the upper limit we derived from the EDGE-CALIFA survey.

Recently, a series of simulations have investigated the ISM properties of galaxies at very high redshifts, and they all invariably found that in such early and compact systems, the ISM
density towards galaxy nuclei is substantial. \citet{trebitsch19} performed a high-resolution, zoom-in simulation of a single massive BH/galaxy system in the early Universe starting from a cosmological hydrodynamical simulation with radiative transfer and accounting for AGN feedback. By $z=5.7$, the mass of their simulated galaxy and BH reached $M_*\sim 2.5\times10^{10}\msun$ and $M_{BH}=1.4\times10^7\msun$, respectively (the maximum AGN luminosity in the simulation is $L_X\sim 10^{43}$\ergs). With a spatial resolution as good as $\sim7$~pc, \citet{trebitsch19} found that, during AGN phases, when the gas is funneled towards the galaxy center, its distribution within 40 pc from the BH hides most of the lines of sight to it. Further, they find that gas distributed up to kpc scales can contribute to the total column density and obscuration at a level that is comparable to the innermost gas. 
Similar results were found by \citet{lupi22} in another high-resolution, zoom-in simulation of an early, more massive system ($M_*=10^{10}\msun$, $M_{BH}\sim6\times10^8\msun$), in which they were able to follow the evolution of gas in different phases. They found that most obscuration is produced by molecular gas distributed over a few hundred pc scales. The covering factor of gas with column densities above $10^{22}$\cm\ is 0.5-0.6 at $z=7$ and rapidly goes to unity at $z\sim8$, when the gas disk of the host galaxy has still to settle and AGN feedback has still to clear the BH surroundings.

Moving from high-resolution, zoom-in to lower-resolution, large-volume simulations, \citet{ni20} investigated the host properties of the QSO population at $z\geq7$ in Blue Tides \citep{feng16}. With a spatial resolution of a few hundreds pc at $z\sim 7$, they found that the host galaxies of high-$z$ AGN feature compact ISM distributions ($\sim1-2$ kpc half-mass radius for the molecular component) whose column density can reach values as high as Compton-thick for massive ($M_*>10^{10}\msun$) systems. In the simulation, the onset of QSO feedback clears a significant portion of the solid angle to galaxy nuclei, yet $\sim90$\% of luminous ($L_X>10^{44}$\ergs) AGN at $z=7$ are expected to be observed along lines of sight with ISM column densities larger than $10^{23}$\cm. This fraction is predicted to be even higher at earlier times. Overall, the results from numerical simulations are consistent with those of our models with $\gamma=2$, and would point to a dominant, yet undiscovered, population of hidden SMBHs at cosmic dawn. 

\subsection{Diffuse hard X-ray nebulae and other diagnostics of ISM-obscured, high-$z$ AGN} 

At high-$z$, the interaction of nuclear photons with the dense ISM clouds may produce detectable diffuse hard X-ray emission across the entire host galaxy. Hard X-ray nebulae around the nuclei of local Seyfert 2 galaxies are now routinely discovered \citep{ma20,jones21}. The first such evidence was presented by \citet{iwasawa03} for NGC 4388. On top of a larger, diffuse soft X-ray emission produced by photo-ionized gas, they also found X-ray emission at $E>$4 keV extending up to a few kpc from the nucleus. Remarkably, they also found extended Fe K$\alpha$ emission at 6.4 keV on the same scales, which points to emission from cold, low-ionization medium. In another local Seyfert 2 galaxy, ESO 428-G014, \citet{fabbiano17} found similar evidence for kpc-scale diffuse hard X-rays, including Fe K$\alpha$ photons. The extended component is responsible for at least 1/4 of the observed 3-8 keV emission, and represents $\sim0.5$\% of the intrinsic X-ray nuclear power. \citet{fabbiano17} estimated that a uniform scattering medium with density of 1 cm$^{-3}$ and \nh$=3.5\times 10^{21}$\cm\ is needed to reproduce the extended hard X-ray luminosity. Alternatively, denser molecular clouds in the galaxy disk may do the job. Our results indicate that the ISM column and volume density of high-$z$ galaxies are far larger than those observed locally, e.g. by 2-3 dex at $z$=3, providing larger reprocessing efficiencies of nuclear X-ray photons. This suggests that prominent hard X-ray nebulae extending on scales comparable with the size of the AGN host galaxies may be observed in high-$z$ systems. Clearly, high-resolution X-ray imaging is needed. The typical half-light radius of massive star forming galaxies at $z=3$ is $\sim2$~kpc \citep{allen17}, which corresponds to a half-light diameter of 0.52''. This is equal to the angular resolution (Half Energy Width, HEW) of \chandra. Therefore, one may hope to marginally resolve hard X-ray and Fe K$\alpha$ emission in distant AGN, either individually if they are sufficiently bright, or through stacking experiments. Recently, \citet{yan21} performed a spectral stacking experiment with \chandra\ on two samples of massive ($M_*>10^{10}\msun$) galaxies at $0.5<z<2$ in the 7Ms CDFS with different star formation rates. They found a stronger Fe K$\alpha$ line in the stacked spectrum of highly star forming galaxies, and interpret this result as possible evidence for enhanced X-ray reflection linked to the cold, star forming gas on galaxy scales. While this is in support of the widespread existence of diffuse X-ray nebulae, confirmation through the analysis of the X-ray spatial extension is needed. Next-generation X-ray missions are not expected to improve on the \chandra\ image quality. {\it Athena}, the large class mission accepted by ESA and planned 
for the 2030s \citep{nandra13}, would reach 5'' HEW at best. The {\it Lynx} \citep{lynx18} and {\it AXIS} \citep{axis18,marchesi20} mission concepts proposed to NASA aim at angular resolutions comparable to that of \chandra\ with $\sim10-30\times$ larger collecting areas. The photon statistics might then compensate for the lack of improved resolution, allowing for an accurate analysis of the source surface density profile in comparison with the instrumental PSF. Ultimately, this may allow for the detection of ISM-induced, diffuse hard X-ray halos in distant AGN beyond their point-like, nuclear emission. 

In addition to diffuse X-ray nebulae, the interaction of nuclear photons with the ISM in high-$z$ QSOs may leave an imprint in the FIR colors of their hosts. For example, based on numerical simulations, \citet{dimascia21} showed that most of the UV radiation in $z\sim 6$ QSOs may be obscured by dust inhomogeneities in their host galaxies on scales of a few hundreds pc. The AGN radiation would heat some of the dust in the ISM to temperatures of 200 K and more, producing warmer MIR-to-FIR colors than in normal star forming galaxies at the same redshifts. In principle, hidden QSOs at $z\sim6$ may be discovered though a combination of e.g. JWST/MIRI data at $\sim25 \mu$m (F2550W filter) and ALMA data in band 6$\div$7 (corresponding to $\sim 4\mu$m and $\sim$120$\div$160$\mu$m rest-frame, respectively). The resolution of JWST/MIRI data ($\sim 0.9$'', i.e. $\sim 5$ kpc at $z=6$) is, however, not sufficient to resolve the dust emission in the QSO host, as the typical half-light radius of the dust continuum is expected to be around 1 kpc (see Fig.~\ref{cfr_z6qso} {\it center}). Therefore, it will not be possible to determine whether warm dust colors extend across the whole host galaxy or are just limited to its nucleus (i.e. to the torus). Wide-field slitless spectroscopy at 3.5~$\mu$m with JWST/NIRCam is expected to be another way to discover hidden QSOs at $z=6$, e.g. through the detection of emission lines at optical rest-frame wavelengths such as H$\alpha$, \nii, \oiii, H$\beta$, which are also key diagnostics for the gas ionization state. Subsequent follow-up observations with JWST/NIRSpec in IFU mode at 0.13'' (700pc) resolution at the same wavelength may probe the morphology of the line emission in the most extended systems, allowing for spatially resolved maps of the ISM ionization state. This would reveal whether the ionizations maps are patchy and irregular, as it may be expected from large-scale obscuration by ISM clouds, or instead show a classic bi-conical structure as in local AGN obscured by a pc-scale torus \citep{maksym17,venturi18}. 

From the ground, near-IR integral field spectrographs assisted by adaptive optics (AO) modules may also provide spatially-resolved maps of the hosts of ISM-obscured, high-$z$ QSOs.  At $z=6$, near-IR spectroscopy, e.g. in the $\sim 1-2.4 \mu$m range, samples emission lines at UV-rest wavelengths, such as \civ, \ciii, \heii, \nev. UV-line ratios have been demonstrated to be a valuable alternative to standard optical diagnostic diagrams to probe AGN activity at high redshift \citep{feltre16,nakajima18,mignoli19}. The AO-assisted near-IR ERIS-SPIFFIER spectrograph \citep{davies18}, expected to be operational at the VLT in 2022, will be able to deliver UV-line ratio maps spatially resolved on scales of a few hundreds pc at $z=6$. In the future, HARMONI \citep{thatte21} at the ELT will provide similar maps at the same redshifts, but with $\sim$60 pc resolution.

Finally, FIR emission lines are another way to probe hidden high-$z$ quasars and determine the distribution and the physical conditions of the cold ISM in their hosts. \citet{vallini19} used a semi-analytic model to investigate the effects of AGN irradiation on the physics of giant molecular clouds distributed in the host ISM. They considered different system geometries, including one featuring a small-scale torus and ISM clouds distributed in a galaxy disk similar to that used in this work. In general, they found that the CO spectral line energy distribution (CO SLED) -- i.e. the CO line luminosity as a function of rotational quantum number $J$ --  is strongly affected by the X-ray irradiation from the AGN, as X-ray photons penetrate deeper into molecular clouds heating their cores to higher temperatures and increasing the luminosity of CO transitions with $J\gtrsim5$. 
The excitation of high-$J$ CO lines increases with X-ray luminosity (and with the system compactness). For instance, CO(7-6)/\cii\ line luminosity ratios as high as $\sim 0.1$ can only be observed in systems hosting QSOs with $L_X\gtrsim 10^{45}$\ergs. ALMA observations at very high resolution (e.g. 0.03'', 200pc at $z=6$, see e.g. \citealt{walter22}) may then track the spatial variations of these line ratios and test how far from the galaxy center the QSO radiation is being absorbed by giant molecular clouds. 

\section{Conclusions}\label{conclusions}

We combined samples of massive ($M_*>10^{10}\;M_{\odot}$) galaxies observed with ALMA to measure the cosmic evolution of the inter-stellar medium (ISM) density and assess
its role in hiding supermassive black holes at early epochs. We considered ALMA-detected objects in the ASPECS and ALPINE large programs, as well as in individual observations of $z\sim 6$ QSO hosts. We then corrected for selection effects and derived the ISM density evolution for the whole population of massive galaxies. We also developed
a simple analytic model for the ISM cloud distribution to estimate the covering factor of galaxy nuclei due to the large-scale ISM in addition to that of pc-scale, circumnuclear material.
Our main results are as follows:

$\bullet$ We used different tracers for the total ISM mass (dust continuum and \cii\ emission) and different proxies for its spatial extent (UV/optical stellar size, \cii\ size) to derive
the ISM surface density $\Sigma_{gas}$ in our galaxy samples. By assuming that most galaxies are disks with exponential ISM density profiles, we estimated that the average column density seen at galaxy centers is \nhism$\sim2\times\Sigma_{gas}$ (Eq.~\ref{nhave}).

$\bullet$ We found that the median ISM column density evolves as $\sim(1+z)^{3.3}$ over the redshift range $z\sim 0-6$ (Fig.~\ref{nhz}). This means that the ISM obscuration towards the nucleus of a $z>3$ galaxy is typically $>100$ times larger than in local systems. At $z\gtrsim6$, the ISM may even be Compton-thick.

$\bullet$ In massive galaxies the ISM is metal rich, and, at $z\gtrsim2$, its median column density is similar to what is typically measured in AGN X-ray spectra (Fig.~\ref{nhzx}). This suggests that the metal content in the ISM of distant galaxies is large enough to absorb significantly the X-ray radiation from the galaxy nucleus.

$\bullet$ Based on the size, mass, and surface density distributions observed for molecular clouds in the Milky Way, we built an analytic model to describe the covering factor towards galaxy nuclei of ISM clouds with different column densities. The model allows for an evolution in the characteristic cloud surface density with redshift, $\Sigma_{c,*}\propto(1+z)^\gamma$. We found that the rate at which the fraction of ISM-obscured nuclei (i.e. the cloud covering factor) increases with redshift anti-correlates with $\gamma$ (Fig.~\ref{fobscfig}): in fact, if $\Sigma_{c,*}$ increases steeply with redshift, then less clouds are needed to match the total, volume-averaged ISM column density observed at high redshifts.

$\bullet$ We included in our galaxy-scale, ISM-obscuration model the contribution from a pc-scale torus, assuming its covering factor decreases with increasing AGN luminosity as observed locally.
We found that, for $\gamma=2$, such a model successfully reproduces the increase of the obscured AGN fraction with redshift that is commonly observed in deep X-ray surveys, both when different absorption thresholds (10$^{22}$, 10$^{23}$~\cm) and AGN X-ray luminosities (10$^{42.5}$, 10$^{44}$~\ergs) are considered (Figs.~\ref{fobsc2_data},\ref{fobsc2_lowlum}). 

$\bullet$ Similarly to what is found by recent numerical simulations, our results suggest that 80-90\% of supermassive black holes in the early Universe ($z>6-8$) are hidden to our view, largely because of the ISM in their hosts. This would prevent standard UV-rest color selection for most SMBHs growing at early epochs.

$\bullet$ We discussed viable methods to probe hidden SMBHs in the early Universe and test whether they are obscured by their host ISM. Spatially resolved maps of emission line ratios at rest-UV, optical, and FIR wavelengths are promising tools that rely on data from e.g. the VLT (in the future E-ELT), JWST, and ALMA, respectively. In addition, next-generation X-ray missions with sub-arcsec resolution may resolve the diffuse, hard X-ray glow expected from the interaction of central AGN photons with the global galaxy-scale ISM.

\begin{acknowledgements}
We acknowledge financial contribution from the agreement ASI-INAF n. 2017-14-H.O. KI acknowledges support by the Spanish MCINN under grant PID2019-105510GB-C33/AEI/10.13039/501100011033 and ``Unit of excellence Mar\'ia de Maeztu 2020-2023'' awarded to ICCUB (CEX2019-000918-M). SM acknowledges funding from the INAF ``Progetti di Ricerca di Rilevante Interesse Nazionale'' (PRIN), Bando 2019 (project: ``Piercing through the clouds: a multiwavelength study of obscured accretion in nearby supermassive black holes''). We finally thank the referee for their useful and insightful comments.

\end{acknowledgements}

\bibliographystyle{aa} 

\bibliography{/Users/gilli/protocluster/biblio}

\end{document}